\documentclass{IEEEoj}
\usepackage{cite}
\usepackage{graphicx}
\usepackage{textcomp}
\usepackage{xcolor}
\usepackage{blindtext}
\usepackage{booktabs}
\usepackage{lipsum}
\usepackage{soul}
\usepackage{adjustbox}
\usepackage[normalem]{ulem}
\useunder{\uline}{\ul}{}
\usepackage{adjustbox,lipsum}
\usepackage{booktabs}
\usepackage{array}
\usepackage[normalem]{ulem}
\usepackage[utf8]{inputenc}
\usepackage[english]{babel}
\usepackage{dirtytalk}
\usepackage{longtable}            
\usepackage{rotating}
\usepackage{multirow}
\usepackage{amsmath} 
\usepackage{hyperref}
\usepackage{comment}
\hypersetup{
    colorlinks=true,
    linkcolor=blue,
    filecolor=magenta,      
    urlcolor=cyan,
}
\usepackage{lscape}
\usepackage{booktabs}
\usepackage[graphicx]{realboxes}
\usepackage[utf8]{inputenc}
\usepackage{pifont}% http://ctan.org/pkg/pifont
\usepackage{graphicx}
\usepackage{svg}
\usepackage[utf8]{inputenc}
\usepackage{adjustbox}
\usepackage{booktabs}
\usepackage{subcaption} % Required for subfigures
\usepackage{csquotes}
\usepackage{balance}
\usepackage{algpseudocode}
\usepackage{algorithmicx}
\usepackage[ruled,lined,linesnumbered]{algorithm2e}
\usepackage{threeparttable}

\usepackage{textcomp} % Include this package

\IEEEoverridecommandlockouts

\newcommand{\JM}[1]{\textcolor{red}{[JM: #1]}} % Comments by JM in red
\newcommand{\FJ}[1]{\textcolor{blue}{[FJ: #1]}}  % Comments by FJ in blue

\newcommand{\contract}[1]{\textit{#1}}    
\newcommand{\functionsc}[1]{\texttt{#1}}

\usepackage{textcomp}
\def\BibTeX{{\rm B\kern-.05em{\sc i\kern-.025em b}\kern-.08em
    T\kern-.1667em\lower.7ex\hbox{E}\kern-.125emX}}
\AtBeginDocument{\definecolor{ojcolor}{cmyk}{0.93,0.59,0.15,0.02}}
\def\OJlogo{\vspace{-4pt}\hskip-4pt\includegraphics[height=18pt]{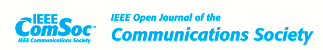}}
\begin{document}
\receiveddate{XX Month, XXXX}
\reviseddate{XX Month, XXXX}
\accepteddate{XX Month, XXXX}
\publisheddate{XX Month, XXXX}
\currentdate{11 January, 2024}
\doiinfo{OJCOMS.2024.011100}

\title{An Empirical Smart Contracts Latency Analysis on Ethereum Blockchain for Trustworthy Inter-Provider Agreements}

\author{Farhana Javed and Josep Mangues-Bafalluy%
\thanks{F. Javed and J. Mangues-Bafalluy are with Services as Networks (SaS), Centre Tecnològic de Telecomunicacions de Catalunya (CTTC/CERCA), Castelldefels, Spain (Email: \{farhana.javed, josep.mangues\}@cttc.es).}
}

\corresp{Corresponding author: Farhana Javed (e-mail: farhana.javed@cttc.es).}

%\authornote{This work was supported by the Natural Sciences and Engineering Research Council (NSERC) of Canada.}
%\markboth{Preparation of Papers for IEEE OPEN JOURNALS}{Author \textit{et al.}}

\begin{abstract}
As 6G networks evolve, \emph{inter-provider agreements} become crucial for dynamic resource sharing and network slicing across multiple domains, requiring on-demand capacity provisioning while enabling trustworthy interaction among diverse operators. To address these challenges, we propose a blockchain-based Decentralized Application (DApp) on Ethereum that introduces four smart contracts—organized into a \emph{Preliminary Agreement Phase} and an \emph{Enforcement Phase}—and measures their gas usage, thereby establishing an open marketplace where service providers can list, lease, and enforce resource sharing. We present an empirical evaluation of how gas price, block size, and transaction count affect transaction processing time on the live Sepolia Ethereum testnet in a realistic setting, focusing on these distinct smart-contract phases with varying computational complexities. We first examine transaction latency as the number of users (batch size) increases, observing median latencies from 12.5 s to 23.9 s in the \emph{Preliminary Agreement Phase} and 10.9 s to 24.7 s in the \emph{Enforcement Phase}. Building on these initial measurements, we perform a comprehensive Kruskal–Wallis test (\( p < 0.001 \)) to compare latency distributions across quintiles of \emph{gas price}, \emph{block size}, and \emph{transaction count}. The post-hoc analyses reveal that high-volume blocks overshadow fee variations when transaction logic is more complex (effect sizes up to 0.43), whereas gas price exerts a stronger influence when the computation is lighter (effect sizes up to 0.36). Overall, 86\% of transactions finalize within 30 seconds, underscoring that while designing decentralized applications, there must be a balance between contract complexity and fee strategies. The implementation of this work is publicly accessible \href{https://github.com/farhanajaved/InP_DApp_Analysis.git}{online}.

\end{abstract}


\maketitle

\section{Introduction}
\IEEEPARstart{T}{he} rapid evolution of network technologies—exemplified by Network Function Virtualization (NFV)—has greatly expanded the benefits of virtualization. By running network functions on general-purpose hardware, NFV enables innovative services while improving the utilization of existing infrastructure \cite{ETSImultidomain}. However, to handle sudden spikes in demand, service providers frequently engage in varying degrees of over-provisioning, which can lead to operational costs when surplus network resources remain underutilized \cite{antevski2022federation}.

At the same time, modern telecommunications ecosystems are no longer dominated by a single operator. Instead, they comprise numerous stakeholders responsible for deploying, operating, and delivering services. This shift has been accelerated by Fifth Generation (5G) mobile networks, which lower barriers to entry for smaller operators by allowing them to lease virtualized resources rather than own physical infrastructure \cite{afraz20205g}. This flexibility empowers diverse providers—large and small alike—to dynamically offer services to broad or niche customer segments.

The rise of Ultra-Reliable Low Latency Communication (URLLC) and Enhanced Mobile Broadband (eMMB) services illustrates the mounting pressure on operators to find balanced, profitable strategies for resource allocation. \textit{Inter-provider agreements} represents one such approach, enabling operators to consume or lease network services and resources from external administrative domains \cite{javed2022blockchain,ETSIGSPDL024}. When a telecommunications provider holds surplus capacity, it can temporarily lease that capacity to others, reducing operational expenses while offering option to resource-consuming entities. In this way, \textit{inter-provider agreements} emerge, supported by an open marketplace for resource sharing among service providers \cite{javed2022blockchainauction,javed2022blockchainPIMRC}. These benefits have made telecommunications operators and service providers more receptive to the idea of inter-provider resource sharing and forming B2B agreements, though concerns persist regarding complexity, trust, and real-world feasibility. The value of these agreements is evident in scenarios requiring specialized geographic coverage—such as large-scale sporting events—and in situations marked by abrupt surges in user demand. Yet, as industry structures evolve, \emph{open marketplaces} must allow smaller players to participate and thrive, necessitating robust trust mechanisms to ensure fairness and reliability.

In this direction, Blockchain is increasingly regarded as a mechanism for distributing trust across these expanding telecommunications environments \cite{javed2022distributed,ETSI2020Applications}. As ownership models become more fragmented—encompassing diverse, smaller entities—the limitations of a single, centralized regulatory body become more apparent. By employing Distributed Ledger Technologies (DLTs), such as blockchain, telecommunications stakeholders can mitigate single points of failure and establish trust without relying on one overarching authority \cite{faisal2023design}. A distributed approach seeks to balance these extremes by leveraging blockchain-based DLTs. Through its inherent transparency and immutability, blockchain can serve as a trustworthy platform for negotiating and executing inter-provider agreements without relying on an intermediary \cite{javed2024blockchain}. 

Nonetheless, implementing a decentralized marketplace for multi-operator resource leasing introduces additional complexities\cite{ETSI2021SmartContracts}. Chief among these is the smart contract execution time associated with blockchain and the parameters that are important to consider while building such a decentralized marketplace and to understand the performance of smart contracts—ranging from initial service agreements to ongoing Quality of Service (QoS) monitoring and Service Level Agreement (SLA) enforcement \cite{javed2023blockchain6GSLA,ETSI2024Settle}.

To address these challenges, this paper proposes a blockchain-based DApp that serves as a marketplace for service providers to share resources and form inter-provider agreements, acting as either \emph{consumers} or \emph{providers}. This marketplace allows participants to list and lease services, enabling direct interaction between them, while smart contracts automate the establishment of mechanisms for monitoring and enforcing KPIs. These inter-provider agreements are supported by multiple smart contracts that facilitate interactions across distinct phases, with their performance analyzed on the live Sepolia Ethereum testnet—a simulated environment mirroring the Ethereum mainnet. To the best of our knowledge, this work is the first to gather real-time \emph{Sepolia} testnet parameters (e.g., block size, gas price) and systematically evaluate their relationships with key smart contract's execution times in the context of inter-provider agreements.

The work contributes to the development and analysis of smart contracts within decentralized marketplaces through three key aspects: (i) a two-phase inter-provider marketplace consisting of four distinct smart contracts—\contract{AddService.sol}, \contract{SelectService.sol}, \contract{RegisterBreach.sol}, and \contract{CalculatePenalty.sol}—tailored to automate critical phases of inter-provider agreements, i.e., the \emph{Preliminary Agreement Phase} and \emph{Enforcement Phase}; (ii) an empirical evaluation of the key smart contracts’ functions on the Sepolia testnet, with experiments measuring transaction latency across batch sizes ranging from 2 to 50; and (iii) a statistical analysis using Kruskal-Wallis and Dunn’s tests to quantify how blockchain parameters (block size, gas price, transaction count) influence transaction latency, thereby providing insights for optimizing blockchain-based smart contracts’ operations, particularly those presented in this use case (inter-provider agreements).

This paper is organized as follows. Section \ref{sec2} presents the related work, highlighting prior approaches and research gaps. Section \ref{sec3} then delves into the background of inter-party agreements and blockchain principles, establishing the context for our study. Section \ref{parameters} outlines the core concepts underpinning our smart-contract analysis, setting the stage for the technical framework. Section \ref{sec5} details the proposed blockchain-based application design, explaining how its components support reliable inter-party agreements. Section \ref{sec6} discusses the deployment strategy and offers a thorough performance evaluation of the implemented solution. Finally, Section \ref{sec7} concludes the paper and identifies avenues for future research.

\section{Related Work and Our Contribution}
\label{sec2}
Blockchain technology has been heralded as a transformative layer of trust in various sectors, with significant attention from standardized organizations within the telecommunications industry. A prime example is the TM Forum, which has been at the forefront of integrating blockchain to standardize and automate contract management and revenue sharing among network operators \cite{tm_forum_blockchain}. Similarly, the CAMARA API initiative focuses on leveraging blockchain to bolster API security and interoperability, thus fostering more secure and efficient digital service ecosystems \cite{camara_blockchain}.

Furthermore, the European Telecommunications Standards Institute (ETSI) has undertaken multiple initiatives that explore blockchain's potential in several critical areas. These include ensuring the integrity of digital identities \cite{etsi_pdl_017}, enhancing the security of IoT networks \cite{etsi_pdl_026}, and facilitating data privacy and consent management across different networks \cite{etsi_pdl_023}. These efforts collectively underline the versatile applications of blockchain in enhancing telecommunications infrastructure.

Beyond standardization initiatives, the broader academic literature on blockchain’s utility in 5G and beyond networks highlights a variety of applications aimed at strengthening security, ensuring privacy, and enabling trustworthy interactions \cite{faisal2022beat,antevski2022federation,baskaran2023role,zeydan2022blockchain}. In the context of wireless networks, blockchain supports essential technical requirements such as security and privacy compliance \cite{augusto2024sok,li2024security,hasan2024blockchain}, enables trustless trading environments \cite{alshahrani2024enabling,tripi2024security}, brokers federated network slices \cite{hafi2024split}, and promotes the integration of private networks with traditional telecommunications operators \cite{han2024lightweight}.

In terms of performance analysis, studies such as \cite{zahir2024performance} assess the efficacy of blockchain within multi-cloud federations by comparing private and public blockchain networks integrated with production-ready orchestration solutions. Additional research by \cite{antevski2023applying} analyzes various consensus mechanisms, providing insights into their efficiencies in different network environments. Moreover, \cite{wilhelmi2022end} and \cite{afraz2023blockchain} address practical considerations like optimizing block size to reduce transaction confirmation latency, as well as the economic implications of deploying blockchain-based solutions in telecommunications. They compare different cost and infrastructure options—on-premises, Infrastructure as a Service (IaaS), and Blockchain as a Service (BaaS)—to select the most suitable blockchain platform.

Our research specifically extends existing guidelines presented by ETSI \cite{ETSI2022SLA,ETSI2024Settle,ETSI2020Applications} to address the challenges of blockchain-enabled decentralized marketplaces, particularly in enhancing and applying inter-provider agreements. In doing so, we focus on the example scenario described by ETSI \cite{ETSI2024Marketplace}, which explores a network slicing resource sharing marketplace. In our framework, blockchain and smart contracts facilitate the trading of network slices—virtual instances of networks on shared physical infrastructure. This decentralized marketplace enables resource or service providers, acting as both consumers and providers, to exchange resources and form administrative relationships through smart contracts. As a result, multiple administrative domains can dynamically participate in an inter-provider marketplace, leveraging the flexibility of a trustless environment.

Within this decentralized marketplace, we propose a comprehensive framework that structures interactions and transactions in multiple phases. The initial phase establishes an open marketplace, where smart contracts handle the preliminary agreement processes. Providers add services to the marketplace, and potential consumers select these services for further negotiation. This approach supports a decentralized inter-provider marketplace, ensuring efficient and transparent exchanges among participants.

Recognizing the necessity of having clearly established Quality of Service (QoS) benchmarks \cite{ETSI2022SLA,ETSI2024Settle} and addressing settlement requirements in the event of penalties, our work implements smart contract-based enforcement mechanisms that monitor Service Level Agreements (SLAs). These mechanisms not only ensure QoS compliance but also manage the settlement processes effectively, assigning penalties and handling transactions. This design is particularly important when dealing with intra-provider agreements that may have varying constraints, as it promotes fairness in penalty execution and compliance across different administrative domains.

Deploying smart contracts on a public blockchain introduces known challenges, notably latency issues \cite{ETSI2021SmartContracts}. We thus undertake a comprehensive analysis of the factors influencing latency in a decentralized marketplace, examining parameters such as gas price, transaction count, and block size. A clear understanding of how these factors scale with increasing user numbers is crucial for designing reliable and efficient smart contract executions within public blockchain environments.

This work presents contributions to the development and analysis of smart contracts within a decentralized marketplace:
\begin{itemize}
    \item  We introduce four distinct smart contracts that support different stages of a blockchain-enabled service lifecycle, from initial registration to breach handling and penalty calculations. This two-phase architecture provides a foundational framework for automating and securing multi-provider network interactions.
    \item The operational performance of these contracts—\contract{AddService.sol}, \contract{SelectService.sol}, \contract{RegisterBreach.sol}, and \contract{CalculatePenalty.sol}—is evaluated on the \emph{Sepolia} testnet, offering insights essential for real-world deployment.
    \item Through extensive experimentation, we investigate how variations in block size, gas price, and transaction count affect transaction latency. Our analysis particularly highlights the scalability of transaction processing as batch sizes (number of administrative domains) increase from 2 to 50.
    \item  We employ Kruskal-Wallis and Dunn’s tests to examine variance in transaction latencies across different network parameters, confirming the significant impact of these factors on performance and providing a quantitative basis for optimizing blockchain operations.
\end{itemize}

%From this analysis on a decentralized marketplace using the Sepolia Ethereum testnet, one can learn how blockchain transaction parameters—specifically, gas price, transaction count, block size, and latency—are influenced by varying batch sizes during smart contract execution. The use of statistical tests like the Kruskal-Wallis (KW) test helps identify whether these parameters exhibit significant differences across defined quantiles, such as varying levels of gas prices or block sizes. Subsequent post-hoc Dunn tests can then elucidate how distinct these quantiles are in practical terms. Ultimately, this analysis aims to offer insights into the scalability and efficiency of blockchain operations, providing key data that can inform optimizations for reducing costs and improving transaction throughput in a blockchain-based marketplace environment. This comprehensive understanding aids stakeholders in making informed decisions about deploying and managing blockchain applications effectively.

\section{Inter-Provider Agreements and Blockchain}
\label{sec3}
This section provides an overview of inter-provider agreements and their existing challenges. Also it explores how blockchain can be utilized within such agreements.

\subsection{Inter-provider agreements: Context and Challenges}
\label{challenges}

The 5Growth project \cite{5GAdvance,20205growth}, part of the EU’s Horizon 2020 initiative, laid the groundwork for multi-domain 5G services by demonstrating how operators could collaborate across different administrative regions to benefit sectors like manufacturing, energy, and transportation. Building on those foundational concepts, proposed \emph{inter-provider agreements} take this collaboration further by integrating separate networks under a common arrangement, enabling consumer domains to tap into specialized capabilities at scale while provider domains monetize previously idle resources \cite{javed2022blockchainPIMRC,javed2022blockchainglobecom}.

By clearly defining terms of collaboration, inter-provider agreements help operators respond rapidly to industrial demands, avoiding costly over-provisioning while supporting seamless service delivery. Existing frameworks like the NFV from ETSI \cite{ETSINFV002} show how standardization and interoperability can strengthen these multi-domain partnerships, and as 5G adoption accelerates, these agreements become essential for harnessing the full potential of service federation across diverse industries.

Inter-provider agreements are crucial for the seamless federation of services and resources across multiple administrative domains, as highlighted in the 5Growth project. However, these agreements present challenges, with the establishment of trust among participating entities being paramount. In an environment where multiple providers and consumers interact, it becomes complex to simultaneously ensure access to services, advertise them, maintain transparency, and enforce SLAs.

These agreements typically progress through various phases, including the setup of agreements and the enforcement of SLAs for lifecycle management, each introducing its own set of challenges. Key among these challenges is ensuring that all parties adhere to the service-level terms, establishing a reliable mechanism for registering and penalizing any breaches. Charging is particularly critical in multi-domain settings, where intricate agreements necessitate meticulous tracking of breaches and accurate calculation of penalties.

\subsection{Applying Blockchain to inter-provider agreements}

A blockchain is recognized as a decentralized ledger that permanently documents all activities within its network. In this paper, we exploit and analyze such feature in Ethereum and the Sepolia testnet. This process involves assembling blocks that encompass a batch of transactions, each marked by unique IDs \cite{vujivcic2018blockchain, wood2014ethereum}. After a block is appended to the chain, any attempt to alter its data necessitates rewriting all subsequent blocks. This architectural design ensures that, once enough blocks are appended, the data in any specific block becomes immutable, thus securing the reliability of recorded information \cite{buterin2014ethereum}.

\emph{Smart contracts}, which are programmable applications running autonomously on blockchain platforms like Ethereum \cite{EthereumSmartContracts}, facilitate the automatic execution of complex agreements. These contracts, often developed in high-level programming languages such as Solidity \cite{ETSI2021SmartContracts}, replicate the logic and stipulations typically found in traditional contracts, thereby simplifying compliance and governance.

The transparency and immutability inherent to blockchain create a trustworthy framework where all transactions and data exchanges are openly verifiable by any participant. This transparency is crucial in environments where collaboration among entities without pre-existing trust is necessary. The immutable nature of the blockchain assures all parties that the records of their transactions are both enduring and accessible to everyone involved, which is vital for accountability and trust in business operations.

In the decentralized inter-provider agreement facilitated by blockchain, platforms enable the listing and negotiation of services and resources, while smart contracts manage the execution and compliance of these agreements \cite{etsi_pdl_026,camara_blockchain}. This setup not only reduces the potential for disputes but also assures the faithful execution of service-level agreements (SLAs).

This blockchain-enabled inter-provider agreement thus offers an open and credible environment where each transaction and adjustment is verifiable and tamper-proof. It allows providers to capitalize on idle resources and enables consumers to access specialized services on demand. The automated functions of smart contracts are crucial in managing negotiations, verifying compliance, and enforcing penalties. Should a provider fail to meet the service thresholds specified in a contract, penalties are promptly imposed and executed automatically, negating the need for centralized arbitration \cite{tm_forum_blockchain}. This methodology also enhances the overall management of service lifecycles, promoting transparency and accountability in multi-provider settings such as those envisioned in 5G and beyond-5G networks. In conclusion, blockchain technology provides a structured approach to addressing the challenges of trust, transparency, and accountability in complex, multi-stakeholder environments.

For this purpose, in the subsequent section \ref{parameters}, we first explain the preparatory concepts of blockchain that will be used throughout the paper, as they are important for understanding the underlying concepts of our framework.

\section{Preparatory Blockchain Concepts and Relevant Network Parameters}
\label{parameters}
This section introduces the fundamental concepts of blockchain and the key network parameters that will be utilized throughout this article.

\begin{figure}[tbp]
\centering
\includegraphics[width=1\columnwidth]{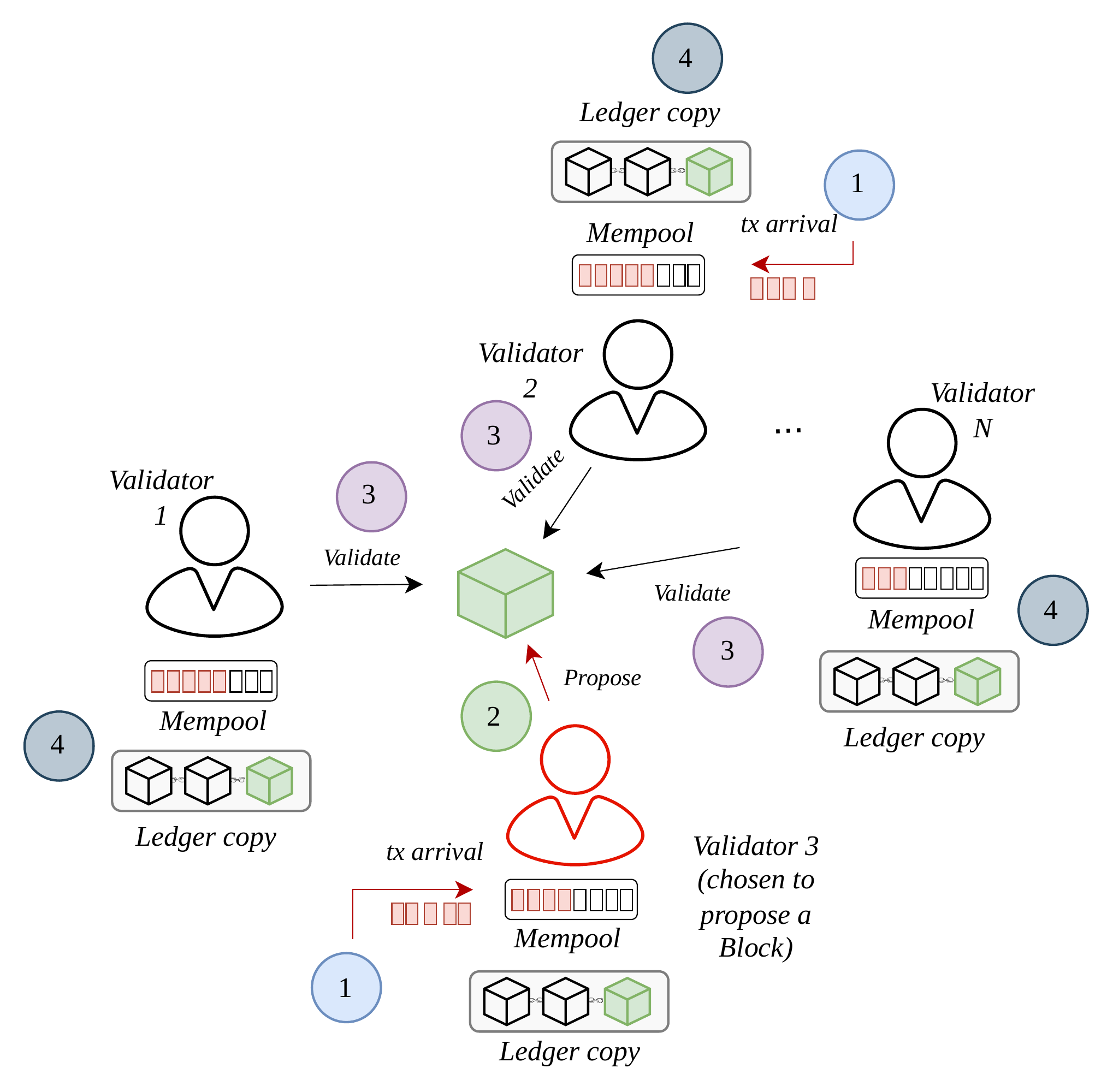}
\caption{PoS-based blockchain operation: (1) Validators receive transactions from users and store them in their transaction pool, (2) a validator is chosen based on their stake to propose a block, (3) other validators validate the proposed block ensuring consensus, (4) once validated, all validators append the new block to their ledger copies.}
\label{fig:PoS}
\vspace{-3mm}
\end{figure}

\subsection{Ethereum: Proof of Stake (PoS): An overview}
Ethereum currently operates under a Proof of Stake (PoS) consensus mechanism, where validators are pivotal in maintaining the network's integrity and security \cite{croman2016scaling,kiayias2017ouroboros}. Validators must stake ETH to participate in block proposal and validation processes. The transaction arrival process begins with validators receiving new transactions from users, which are then stored in their mempool (1) as shown in Figure~\ref{fig:PoS}. The selection of a validator to propose a new block is determined by their staked ETH and a randomized selection mechanism. This validator compiles a block from transactions within their mempool and proposes it to the network (2). Following the proposal, other validators work to verify the validity of the transactions contained in the block, ensuring they comply with Ethereum’s established rules (3). Once consensus is achieved on the proposed block, all validators update their ledger copies to include the new block, extending the blockchain  (4). This consensus mechanism not only secures the network but also updates the blockchain state across all nodes \cite{eyal2016bitcoinng}.

\subsection{Contract Transactions in Ethereum}
In Ethereum transactions are the primary mechanism by which state updates occur on-chain.
There are two broad types of transactions:\emph{User transactions}, which transfer ETH (cryptocurrency of Ethereum) from one externally owned account (EOA) to another.
On the other hand, \textbf{\textit{Contract transactions}} invoke functions in a deployed smart contract. This type of transaction is initiated when an EOA sends a request to execute a function within a smart contract. The transaction encapsulates both the intention and the necessary data to invoke the smart contract's function, which, upon successful validation and acceptance into a block by the network, triggers the execution of the specified function. The execution of a smart contract, facilitated by the Ethereum Virtual Machine (EVM), entails the operational enactment of the bytecode comprising the smart contract's logic. Thus, a contract transaction acts as a mechanism for state change on the blockchain. These may update contract storage, trigger events, or transfer ETH according to the contract’s logic \cite{EthereumTransactions}.

\begin{figure}
    \centering
    \includegraphics[width=1\columnwidth]{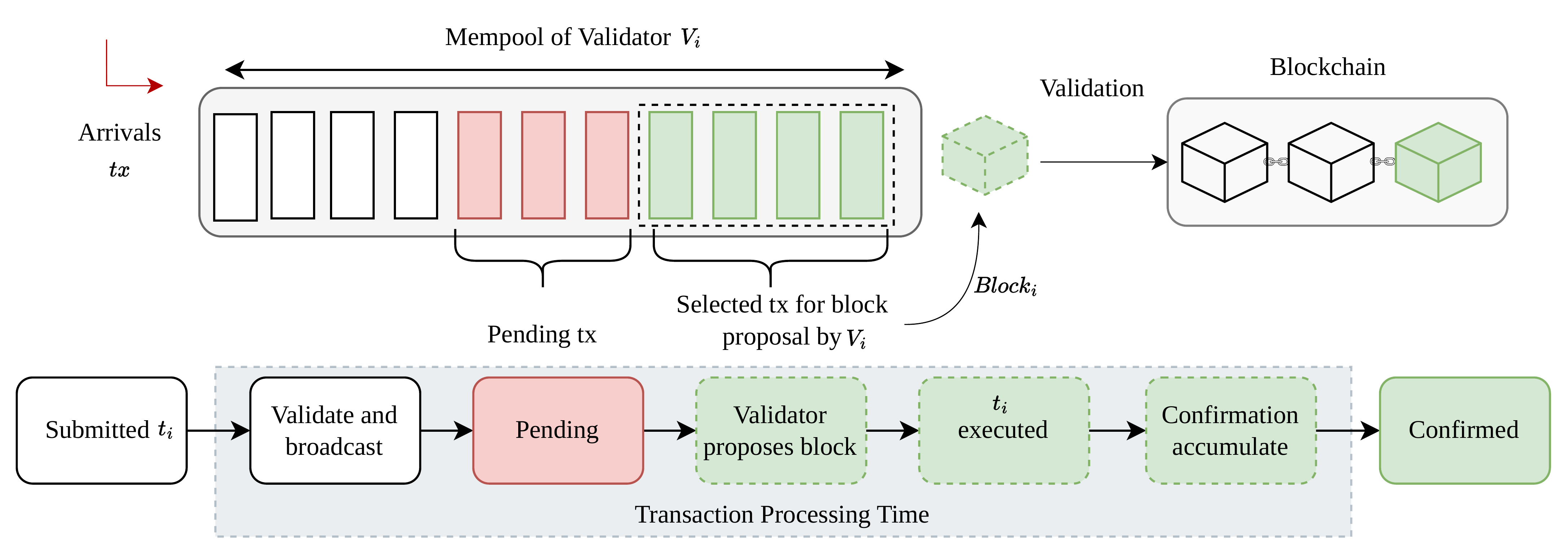}
    \caption{Lifecycle of a Transaction $t_i$ in Ethereum}
    \label{fig:life-cyecle-tx}
\end{figure}

\subsubsection{Lifecycle of a Transaction}
\label{sec:lifecycle-of-a-transaction}

As illustrated in Figure~\ref{fig:life-cyecle-tx}, the lifecycle of a transaction. For example a contract transaction initiated by an externally owned account to execute a function, encapsulating the execution of contract logic, thus modifying the state on the blockchain. A transaction $t_i$ in Ethereum—begins at the moment when a transaction issuer creates and signs $t_i$ using a private key. Upon submission, the issuer’s client or wallet broadcasts the signed transaction to a node in the Blockchain network, which immediately performs basic validation checks.
If these checks pass, the receiving node propagates $t_i$ to other nodes, thereby ensuring it reaches a broad set of validators~\cite{EthereumTransactions}.

Once validated and diffused across the network, the transaction enters the mempool of each validator, where it remains in a \emph{pending} state awaiting block inclusion. Under Ethereum’s PoS consensus, validators (who take the role of block proposers) monitor their mempool for eligible transactions.  When it is a particular validator’s ($V_i$ turn to propose a new block, that validator selects a subset of pending transactions, executes them in deterministic order via the EVM, and packages the resulting state updates into a newly formed block $Block_i$.  The block is then broadcast to the rest of the network, where other validators verify its contents and attest to its correctness~\cite{EthereumProofOfStake,sai2021performance}.

If the block is deemed valid, it is appended to the canonical chain on Sepolia, thereby marking each of its transactions—including $t_i$ as processed on-chain.  At this point, the user can observe that $t_i$ has been “included” in the blockchain, though full confidence in its irreversibility accrues over subsequent blocks.  Each new block referencing the block containing $t_i$ contributes to $t_i$s confirmations, and once enough validators reach consensus on these successive blocks, the chain finalizes. %Under Proof of Stake, finality is achieved through periodic checkpointing; reverting a finalized block requires a prohibitively costly slash of validator stakes.  
Consequently, once these confirmations and checkpoint attestations have accrued, \( t_i \) is regarded as fully confirmed and practically immutable. This final state marks the end of its lifecycle, from initial submission to permanent on-chain inclusion on the Ethereum network~\cite{EthereumTransactions}. Lastly submitting a transaction in Ethereum—whether it is a simple ETH transfer or a complex contract call—incurs a fee. Formally, this fee is computed as: \( \text{Transaction Fee} = \text{Gas Usage} \times \text{Gas Price} \), which is explained \ref{GasUsage} and section \ref{subsec:gas-price} respectively.

\subsection{Gas Usage in Ethereum Transactions}
\label{GasUsage}
The term \emph{Gas Usage} quantifies the computational resources consumed by a transaction’s execution.  
Under the EVM, each opcode (e.g., writing storage, arithmetic, event emission) 
is assigned a certain gas cost~\cite{EthereumVirtualMachine}. For a simple ETH transfer, the gas usage is typically fixed at \(21{,}000\,\text{gas}\). In contrast, \emph{contract transactions}, which involve the execution of functions defined in smart contracts, generally incur higher gas usage. This increase is due to the complexity of the tasks performed, such as multiple function calls, intricate state updates, or event logging. When an EOA initiates a contract transaction, it requests the EVM to execute specified contract logic, leading to varied gas costs depending on the computational intensity of the executed functions \cite{chen2017contracts}.
%Contract transactions, however, tend to consume more gas because they invoke additional opcodes and often modify 
%on-chain state. Consequently, the same gas price can result in a notably higher fee for a contract transaction 
%than for a simple ETH transfer. 
%By adjusting the gas price upward, a user signals higher urgency and can secure faster block inclusion 
%(if network congestion permits).

\subsection{Gas Price in Ethereum}
\label{subsec:gas-price}

The \emph{Gas Price}, denominated in Gwei (where \(1\ \text{Gwei} = 10^{-9}\,\text{ETH}\)), represents the cost per unit of computational effort required to perform transactions or execute smart contracts on Ethereum and similar PoS-based networks. Users effectively \emph{bid} a gas price in order to have their transactions included in the 
next block. Validators, who propose and finalize blocks, typically prioritize higher-paying transactions because they yield greater fee revenue (net of any amounts burned under EIP-1559). Gas prices are determined by supply and demand within the network’s transaction market. Under heavy congestion, users compete for immediate block space, driving gas prices upward. Conversely, when network activity is lower, gas prices tend to fall. This dynamic helps regulate the blockchain’s computational workload by incentivizing validators to include transactions that offer higher fees.

\subsection{Block Size in Ethereum}
\label{subsec:block-size}

In Ethereum, a block is essentially a container for transactions, and its block size (measured in bytes) reflects the total amount of data those transactions occupy. The block size thus determines how much information must be propagated across the network when a new block is broadcast to participating nodes, thereby influencing network latency and overall throughput. Underlying this, Ethereum imposes a block gas limit—often around 30 million gas—that caps the total computational effort of all included transactions. While this limit does not directly translate to the block size in bytes (because computational cost, measured in gas, is not identical to data size), it nonetheless constrains the maximum number and complexity of the transactions that can be included in a single block. Consequently, if a block consumes most of its gas allowance on fewer, highly complex transactions, its byte size may be relatively smaller than a block containing many simpler transactions, or vice versa. In the present study, we focused on the block size itself, along with the number of transactions and the gas price, to evaluate how many transactions were processed concurrently and how large the resulting block was in terms of raw data—factors that affect how quickly the block is disseminated and verified by nodes on the network.

\subsection{Transaction Count in Ethereum}
\label{subsec:transaction-count}
\emph{Transaction Count} denotes the total number of individual transactions that are included within a single block on the Ethereum blockchain. This metric is a critical indicator of the network’s throughput, representing the capacity of the blockchain to process and validate transactions within each block interval. The transaction count is inherently influenced by both the block gas limit and the gas usage of each transaction. Transactions that are straightforward, such as basic Ether (ETH) transfers, typically consume less gas, enabling a greater number of such transactions to be accommodated within a single block. In contrast, more complex transactions—such as those involving smart contract interactions, multiple function calls, or extensive state modifications—require higher gas consumption per transaction. Consequently, blocks containing a higher proportion of these intricate transactions will reach the gas limit with fewer transactions, thereby reducing the overall transaction count per block. This dynamic interplay between transaction complexity and gas consumption directly impacts the network’s efficiency and scalability, as it determines how many users can have their transactions processed in a given timeframe.

Although these three parameters—gas price, block size, and transaction count—are distinct, they collectively influence \textit{transaction latency} (the time from when a user submits a transaction until it is confirmed on-chain). In particular, higher gas prices can incentivize validators to include transactions more quickly, but severe congestion may still delay finalization if the block’s gas limit is saturated \cite{bonneau2015sok,kiayias2017ouroboros}. Conversely, increasing the block size allows more transactions per block but also increases propagation and validation overhead, potentially slowing the network. In addition, the mix of transaction types—from simple ETH transfers to more complex contract calls—directly constrains how many transactions can be packed into each block. Ultimately, each parameter introduces trade-offs in how fast, how many, and which transactions get processed, shaping overall throughput and latency \cite{androulaki2018hyperledger}.

%We present the evaluation of smart contracts as outlined in Section \ref{smart-contracts}, focusing on both phases as presented in \ref{smart-contracts-phase1} and \ref{smart-contracts-phase2}.

\begin{figure*}
    \centering
    \includegraphics[width=1\linewidth]{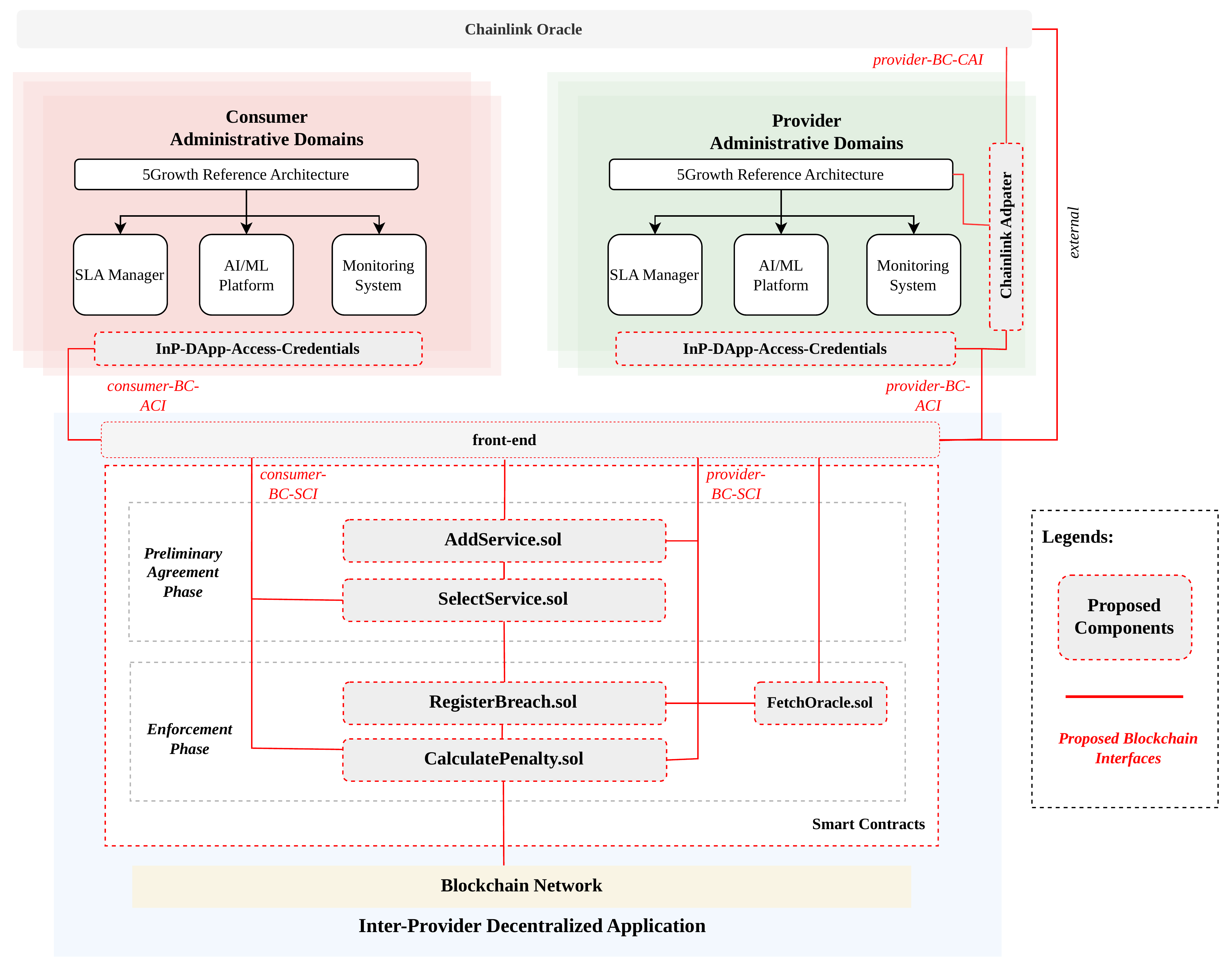}
    \caption{Blockchain-enabled Marketplace for Inter-Provider Agreements: Frameworks and Components and their Interactions}
    \label{fig:framework}
\end{figure*}

\section{Proposed Blockchain Decentralized Application (BC-DApp) for Inter-Provider Agreements}
\label{sec5}

To address the challenges outlined in Section \ref{challenges}, particularly those related to trust, transparency, and enforcement in inter-provider agreements, we propose a blockchain-based marketplace framework structured around two distinct operational phases: (i) the \emph{Preliminary Agreement Phase} and (ii) the \emph{Enforcement Phase}. These phases collectively govern the lifecycle of inter-provider agreements, encompassing the initial advertisement and selection of services, the formalization of commitments, and the ongoing verification of compliance, including the resolution of contractual breaches. 

As depicted in Figure \ref{fig:framework}, our proposed Inter-Provider DApp integrates the blockchain as an additional layer of trust. The proposed framework is divided into two primary Administrative Domains— Consumer and Provider—each incorporating the 5Growth Reference Architecture and blockchain-adapter. The consumer and provider domains expose the following main interfaces (highlighted in red) for inter-provider interactions:
\begin{itemize}
    \item Access-Interface (ACI): These interfaces govern assigned credentials (e.g., Ethereum address and a private key) that define their role (e.g., consumer or provider) within the DApp. The ACI enforces role-based permissions, ensuring ADs can only access the smart contracts relevant to their assigned role.
    \item Smart Contract Interface (SCI): These interfaces define how each administrative domain’s internal components interact with the set of \emph{smart contracts}.
    \item Chainlink Adapter Interface (CAI) as \emph{Provider-BC-CAI}: Provider administrative domain integrates a \emph{Chainlink Adapter} to securely exchange data between off-chain environments and the blockchain.
\end{itemize}

The workflow of our proposed framework commences upon the successful authentication of consumers and providers, each of whom has been assigned an Ethereum address and specific roles. Each registered domain can access the front-end (step 1) using their credentials. Once authenticated, they are granted access to the smart contracts based on their designated roles.

For instance, during the \emph{Preliminary Agreement Phase}, providers initiate the process by listing services using \contract{AddService.sol} (step 2), whereas consumers select from the available services listed (step 3). The conclusion of this phase, marked by the addition and selection of services, seamlessly transitions into the \emph{Enforcement Phase}.

In this phase, key smart contracts such as \contract{RegisterBreach} and \contract{CalculatePenalty} come into play. \contract{RegisterBreach}  enables providers to transmit their KPIs via the \contract{FetchOracle} smart contract (step 4). Upon reaching a predefined threshold of breaches, the \contract{CalculatePenalty} smart contract is activated to assess penalties based on the accumulated number of breaches (step 5).

Below, we provide a detailed discussion of these smart contracts, focusing on their operational complexities and the associated gas consumption.

\begin{table*}[ht]
\centering
\begin{threeparttable}
\caption{Overview of Smart Contract Functions and Interactions}
\label{tab:smart_contracts}
\begin{tabular}{p{3.5cm}|p{5cm}|p{5cm}}
\toprule
\textbf{Smart Contracts} & \textbf{Main Functions and Description} & \textbf{Gas Used} \\ 
\midrule
\contract{AddService.sol} & \functionsc{addService}: Allows each provider to add up to 5 services & 162,500 for the first service, 147,500 for subsequent services \\
\contract{ServiceSelection.sol} & \functionsc{selectService}: Allows consumers to select a service from providers' offerings & Service ID 1: 138,752, Service ID 2: Initially selected (cold access) 155,992, Subsequent: 138,892, Service ID 3: 139,032, Service ID 4: 139,172, Service ID 5: 139,312\\ \midrule
\contract{RegisterBreach.sol} & \functionsc{registerBreach}: Records the occurrence of breaches for the provider & 44,058 (First-Time Execution), 26,958 (Subsequent Executions) \\   
\contract{CalculatePenalty.sol} & \functionsc{calculatePenalty}: Calculates a penalty for a specified user based on the number of breaches they have committed & 49,143 \\ 
\bottomrule
\end{tabular}
\end{threeparttable}
\end{table*}

\subsection{Proposed Smart Contracts and Associated Gas Usage}
\label{smart-contracts}
The following sections provide a detailed examination of the smart contract functionalities underpinning each phase, elucidating their operational mechanisms and evaluating their computational complexity in terms of gas consumption. Table \ref{tab:smart_contracts} presents an overview of key smart contract functions for both phases within a decentralized application as shown in Figure \ref{fig:framework}, detailing their main functionalities, the batch sizes for transactions, and the gas consumption for each operation.

\begin{algorithm}[tp!]
\caption{\emph{Preliminary Agreement Phase} Smart Contracts}
\label{alg:add_select_service}
\footnotesize
\SetKwInOut{Input}{Input}
\SetKwInOut{Output}{Output}
\SetKwProg{Fn}{Function}{}{end}
\SetKwProg{Pn}{Contract}{}{end}

\Pn{AddService}{
    \textbf{State Variables:}\;
    \quad\textit{services}: mapping(uint256 $\rightarrow$ Service)\;
    \quad\textit{providerAddresses}: address[]\;

    \Fn{addService(serviceId, location, cost) \textbf{public}}{
        provider $\gets$ \textit{msg.sender}\;
        \If{provider not in providerAddresses}{
            providerAddresses.append(provider)\;
        }
        services[serviceId] $\gets$ \textbf{new Service} with $(serviceId, location, cost)$\;
        \textbf{emit} \textit{ServiceAdded}(provider, serviceId, location, cost)\;
    }

    \Fn{getService(serviceId) \textbf{public} \textbf{view} \textbf{returns} \textbf{(Service)}}{
        \Return services[serviceId]\;
    }
}

\Pn{SelectService}{
    \textbf{State Variables:}\;
    \quad\textit{addServiceContract}: address\;
    \quad\textit{selections}: array of \textit{Selection}\;

    \Fn{\textbf{Constructor}(addServiceAddress)}{
        \textit{addServiceContract} $\gets$ \textbf{AddService at addServiceAddress}\;
    }

    \Fn{selectService(provider, serviceId) \textbf{public}}{
        service $\gets$ \textit{addServiceContract}.getService(serviceId)\;
        newSelection $\gets$ \textbf{new Selection} with $(msg.sender, provider, serviceId)$\;
        selections.append(newSelection)\;
        \textbf{emit} \textit{ServiceSelected}(msg.sender, provider, serviceId)\;
    }
}
\end{algorithm}

\subsubsection{Preliminary Agreement Phase}
\label{smart-contracts-phase1}
In the \emph{Preliminary Agreement Phase}, as illustrated in Figure \ref{fig:framework}, service offerings are advertised by provider and agreements are formed. This phase is crucial because it establishes the foundation of the inter-provider agreements and includes the following smart contracts as discussed in Section \ref{addition} and \ref{selection}. 

\paragraph{Addition of a Service} 
\label{addition}
In the proposed inter-provider DApp, each provider is permitted to advertise up to $n$ number of services. The smart contract \contract{AddService.sol} enables this functionality through its key function \functionsc{addService}, which allows providers to input details about each service they offer. This mechanism maps each service to its respective provider, ensuring unique identification and accessibility. To further illustrate the \functionsc{addService} function and its impact on gas consumption, we present a simplified algorithm in Algorithm~\ref{alg:add_select_service}.

As discussed in Section \ref{parameters}, \emph{gas costs} inherently accompany all contract transactions—in this case, whenever any provider interacts with \contract{AddService.sol}. We limit our analysis to the addition of up to five services for each provider, striking a balance between capturing detailed insights into gas consumption and conserving resources for our testing.

Gas usage reveals that the first service added incurs higher gas consumption (approximately 162,500 units for Service ID 1) due to the \emph{initialization} of the service array for each provider. This process is resource-intensive as the smart contract transitions "cold" storage to an initialized state by creating a new array within the provider’s mapping—a process demanding additional computational steps \cite{wood2014ethereum,chen2017contracts}. Specifically, the mapping \textit{providerServices[provider]} is initialized during this first transaction (Line 5 in Algorithm~\ref{alg:add_select_service}), incurring higher gas costs due to storage allocation and initialization \cite{decentralizedstorage2024}. Each transaction initiated by a provider using the \textit{addService} function maps the new service to that provider within the existing array, establishing a unique connection and impacting gas costs due to the computational overhead associated with initializing state variables.

For subsequent additions (Service IDs 2 to 5), the gas usage decreases to approximately 147,500 units and remains consistent. This reduction is attributed to the service array already being initialized with the first service. Thus, subsequent additions involve accessing "warm" storage, which is less resource-intensive as it does not require further memory allocation or state initialization \cite{decentralizedstorage2024}. The function merely appends the new service index to the existing array (Line 7 in Algorithm~\ref{alg:add_select_service}), resulting in lower and consistent gas usage for these services, as the storage structures are already in place and only minimal state changes are required.

\paragraph{Selection of a Service}
\label{selection}
In the proposed inter-provider DApp, the service selection phase enables each consumer to choose a service from the list offered by all providers. The smart contract \contract{ServiceSelection.sol} facilitates this through its \functionsc{selectService} function, crucial for the DApp’s operation as it involves write operations that update the blockchain state by mapping consumer choices to specific service IDs.

The \functionsc{selectService} function, detailed in Algorithm~\ref{alg:add_select_service}, operates by first verifying the existence of the chosen provider and service. It accesses the \texttt{providers} mapping to retrieve the provider's details and checks if the specified service ID corresponds to a service offered by that provider. Upon validating these conditions, the function records the selection by appending a new \texttt{Selection} struct to the \texttt{selections} array, capturing the consumer's address, the provider's address, and the selected service ID.

As discussed previously in the context of the \contract{AddService.sol} function, initializing storage structures when selecting a service for the first time incurs higher gas costs, as outlined in Algorithm~\ref{alg:add_select_service}. This initial selection involves "cold" storage access, similar to the first service addition. For example, selecting Service ID 2 for the first time uses approximately 155,992 units due to the overhead of initializing the mapping and storing the selection data.

Subsequent selections exhibit lower gas usage, which varies slightly depending on the service’s position within the provider's list, as the deeper service positions require more SLOAD \cite{EtherVM} operations for access. This is depicted in Algorithm~\ref{alg:add_select_service}, where accessing services further down the list incurs additional costs. For instance, selecting Service ID 1 results in gas usage of about 138,752 units, escalating incrementally to approximately 139,312 units for Service ID 5 due to increased storage reads.

%In summary, the gas usage during the service selection process is directly influenced by both the initialization of storage structures and the position of the service within the provider's list. The initial higher gas cost for first-time selections is attributed to storage allocation and state initialization. Subsequent selections benefit from the already initialized structures, resulting in lower gas costs.

%By analyzing these patterns, developers can better estimate transaction costs and design smart contracts that are both cost-effective and scalable. This ensures a more efficient and user-friendly DApp, ultimately benefiting all participants in the blockchain network.

\subsubsection{Enforcement Phase}
\label{smart-contracts-phase2}
This phase is dedicated to monitoring compliance with service agreements and managing breaches. It ensures that providers fulfill their contractual obligations and establishes mechanisms through smart contracts to address any violations effectively. This phase plays a crucial role in maintaining the integrity and trustworthiness of the service provision within the decentralized application.
\begin{algorithm}[t!]
\caption{Breach Registration and Penalty Calculation Process}
\label{alg:breach_penalty_process}
\footnotesize
\SetKwInOut{Input}{Input}
\SetKwInOut{Output}{Output}
\SetKwProg{Pn}{Contract}{}{end}
\SetKwProg{Fn}{Function}{}{end}

\Pn{RegisterBreach}{
    \textbf{State Variables:} \\
    \quad \texttt{breaches}: mapping(address $\rightarrow$ uint256)\\[6pt]

    \Fn{registerBreach(numBreaches) \textbf{public}}{
        sender $\gets$ msg.sender\\
        \texttt{breaches}[sender] $\gets$ \texttt{breaches}[sender] + numBreaches\\
        \textbf{emit} \texttt{BreachRegistered}(sender, numBreaches)\\
    }
}

\Pn{CalculatePenalty}{
    \textbf{State Variables:} \\
    \quad \texttt{breachContract}: RegisterBreach\\
    \quad \texttt{penalties}: mapping(address $\rightarrow$ uint256)\\[6pt]

    \Fn{\textbf{constructor}(registerBreachAddress) \textbf{public}}{
        \texttt{breachContract} $\gets$ RegisterBreach(registerBreachAddress)\\
    }

    \Fn{calculatePenalty(user) \textbf{public returns} (uint256)}{
        breachCount $\gets$ \texttt{breachContract.breaches}(user)\\
        fidelityFee $\gets$ 1\\
        penalty $\gets$ fidelityFee $\times$ breachCount\\
        \texttt{penalties}[user] $\gets$ penalty\\
        \textbf{emit} \texttt{PenaltyCalculated}(user, penalty)\\
        \Return penalty\\
    }
}
\end{algorithm}

\paragraph{Registration of a Breach}
\label{breach}
The \contract{RegisterBreach.sol} smart contract has one main function, \functionsc{registerBreach}, that registers breaches as they occur. We present an Algorithm in \ref{alg:breach_penalty_process}. The \functionsc{registerBreach} function in the \contract{RegisterBreach.sol} contract is essential for managing violations of SLAs by service providers. 

To accurately register breaches by transferring off-chain performance metrics on-chain, this smart contract is integrated with \contract{FetchOracle.sol}. In our system, service providers utilize the Chainlink Adapter to transmit their KPIs to the Chainlink decentralized oracle network. Subsequently, the Chainlink oracle forwards this data to a mock smart contract named \contract{FetchOracle.sol}. This contract acts as an intermediary, retrieving the KPI data and making it available to our \textit{RegisterBreach.sol} smart contract. This integration with Chainlink ensures that off-chain performance metrics are securely and accurately transferred on-chain, thereby preserving data integrity and enhancing trustworthiness.

This function operates by incrementing the count of breaches recorded against a service provider whenever a violation is reported via \contract{FetchOracle.sol}. Each provider has an associated breach count, which is stored within a mapping called \texttt{breaches} (see lines 1 and 7 of \ref{alg:breach_penalty_process}) that links provider identities to their respective breach histories. To ensure that penalties are not continuously incurred and to maintain gas efficiency, the function also integrates a cap on the number of breaches a provider can accrue, set by \textit{maxBreach}, at three. Once this cap is reached, the \textit{calculatePenalty} function is automatically triggered, calculating the penalties based on the total number of breaches recorded.

From a gas consumption perspective, the \texttt{registerBreach} operation exhibits similar characteristics as explained above in Section \ref{addition}. It incurs a 'cold' storage cost the first time it is executed for each provider. This results in a higher initial gas usage—approximately 44,058 units. Subsequent executions for the same provider benefit from 'warm' storage operations, reducing the gas cost to around 26,958 units.

\paragraph{Calculation of Penalty}
\label{penalty}
The \textit{calculatePenalty} function, within the \textit{CalculatePenalty.sol} contract, is designed to compute the financial penalties for breaches by service providers using a linear formula. Upon activation, after the breach limit is reached via the \textit{registerBreach} function, \textit{calculatePenalty} retrieves the current breach count from the \texttt{breaches} mapping (see Algorithm~\ref{alg:breach_penalty_process}). It then applies a computation: \(\texttt{penalty} = \texttt{fidelityFee} \times \texttt{breachCount}\)
as illustrated on line~20 of Algorithm~\ref{alg:breach_penalty_process}. 
Here, \(\texttt{fidelityFee}\) can be any constant set by the contract (e.g., 10), and \(\texttt{breachCount}\) denotes the total number of recorded infractions. After calculating the penalty, this amount is recorded in a separate mapping called \texttt{penalties} (line~21), associating each provider with their corresponding financial penalty. Finally, an event is emitted (line~22) to signal the penalty assessment and completion of \emph{Enforcement Phase}.

Gas usage for \texttt{calculatePenalty} costs approximately 49,134 units. This cost reflects several operations, including reading breach counts from storage (line~18), performing arithmetic (line~20), writing the computed penalty to storage (line~21), and emitting a penalty event (line~22). By invoking \texttt{calculatePenalty} only after the maximum breach threshold is reached, the contract reduces unnecessary repetitive calculations, achieving a more gas-efficient mechanism for enforcing penalties.

%This automation and integration between the two functions streamline the process of breach registration and penalty imposition, optimizing the contract's operation and gas usage within the blockchain environment. In this simplified approach, the cost of performing the arithmetic on-chain is minimal compared to the overall write operations. Nonetheless, some implementations still choose to do part of the calculation or trigger logic off-chain. Regardless, the on-chain cost primarily comes from the \texttt{calculatePenalty} function call itself and subsequent storage writes, not from the linear penalty math.

\section{Deployment and Evaluation} 
\label{sec6}

This section details our methodology for simulating and evaluating Ethereum-based smart contracts, as introduced in \ref{sec3}. The environment comprises Kubuntu 22.04.3 LTS with 32\,GB DDR4 RAM, Hardhat 2.22.4 for development, and Solidity v0.8.0 for creating smart contracts, summarized in Table~\ref{tab:system_setup}. Hardhat was chosen for its straightforward development workflow, strong plugin ecosystem, all of which streamline testing and deployment of Ethereum smart contracts. Alchemy, a blockchain infrastructure provider, was used to interface between the Hardhat framework and the Sepolia testnet. It abstracts low-level network complexities, manages transaction routing, and ensures stable connectivity, which is critical for reproducible experiments and accurate gas estimation under dynamic network conditions.

We investigate how transaction latency influences smart contract functionality for inter-provider agreements. Each transaction, identifiable by a unique hash, is monitored to measure and log this latency while scaling the number of interacting Externally Owned Accounts (EOAs). We also capture key network parameters to understand performance under varying conditions.

Our analysis proceeds in two parts. First, we examine how transaction latency changes when batch sizes grow from 2 to 50 across different smart contract functions. Second, we perform a detailed statistical analysis on parameters such as gas price, transaction count, and block size to uncover their impact on latency.

\begin{table}[t]
\centering
\begin{threeparttable}
\caption{Simulation system setup for inter-provider agreements on blockchain}
\label{tab:system_setup}
    \begin{tabular}{p{3cm} p{5cm}}
        \textbf{Component} & \textbf{Specification} \\ \toprule
        Operating system & Kubuntu 22.04.3 LTS\\ \midrule
        Memory (RAM) & 32 GB DDR4 \\ \midrule
        Blockchain testnet & \href{https://sepolia.etherscan.io/}{Sepolia testnet}\tnote{1} \\ \midrule
        Node service & \href{https://www.alchemy.com/}{Alchemy}\tnote{2} \\ \midrule
        Development framework & \href{https://hardhat.org/}{Hardhat} 2.22.4\tnote{3} \\ \midrule
        Smart contract language & \href{https://soliditylang.org/}{Solidity} v0.8.0\tnote{4} \\ \midrule
        Number of accounts & 100 Externally Owned Accounts (EOAs) and their private keys\tnote{5}  \\ \midrule
        Iteration Rounds & 10 \\ \bottomrule
    \end{tabular}
    \begin{tablenotes}
    \item[1] A recommended default testnet for application development.
    \item[2] A blockchain node service providing APIs and infrastructure for reliable network connectivity and gas estimation.
    \item[3] A development environment for Ethereum.
    \item[4] Programming language designed for developing smart contracts that run on Ethereum.
    \item[5] An Ethereum account is an entity with an ether (ETH) balance that can send transactions.
    \end{tablenotes}
\end{threeparttable}
\end{table}

\subsubsection{Transaction Latency}
\label{define-latency}

Latency is a pivotal performance metric in blockchain systems, reflecting the interval between a transaction’s submission and its confirmation on-chain \cite{bonneau2015sok}. We quantify it by subtracting the submission timestamp ($T_{\text{submitted}}$) from the block’s timestamp ($T_{\text{confirmed}}$) that includes the transaction:
$L = T_{\text{confirmed}} - T_{\text{submitted}}$.

By leveraging the Sepolia testnet, we measure real-time transaction processing in a live environment. Our objective is to identify factors—such as gas price (cost per gas unit), block size (block capacity), and transaction count—that  affect latency.

To investigate these variables, we send concurrent transactions from multiple accounts to various smart contract functions, systematically adjusting batch sizes (the number of consumers or providers). We collect transaction data via \href{https://sepolia.etherscan.io/}{sepolia.etherscan}, gas price (in gwei), block size (in KB), and transaction count are each divided into quintiles (Q1–Q5), representing 20\% intervals in their data distributions.

Statistical tests are then applied to analyze these distributions. We begin with the Kruskal–Wallis test, a non-parametric method assessing whether medians differ across multiple groups. If significant differences arise, we use Dunn’s post-hoc test (with Bonferroni correction) to identify which pairs of quintiles differ significantly, and then calculate Cliff’s Delta to measure effect sizes. Established thresholds categorize effect sizes as negligible (\(|\delta| \leq 0.147\)), small (\(0.147 < |\delta| \leq 0.33\)), medium (\(0.33 < |\delta| \leq 0.474\)), or large (\(|\delta| > 0.474\)). Through this approach, we quantify how each parameter influences network latency and assess the practical significance of observed differences.

\subsection{Transaction Latency and Batch Size}
\subsubsection{\functionsc{addService} and \functionsc{selectService}}
\label{sec-phase1-latency}
\begin{figure}[ht!]
    \centering
    \includegraphics[width=1\linewidth]{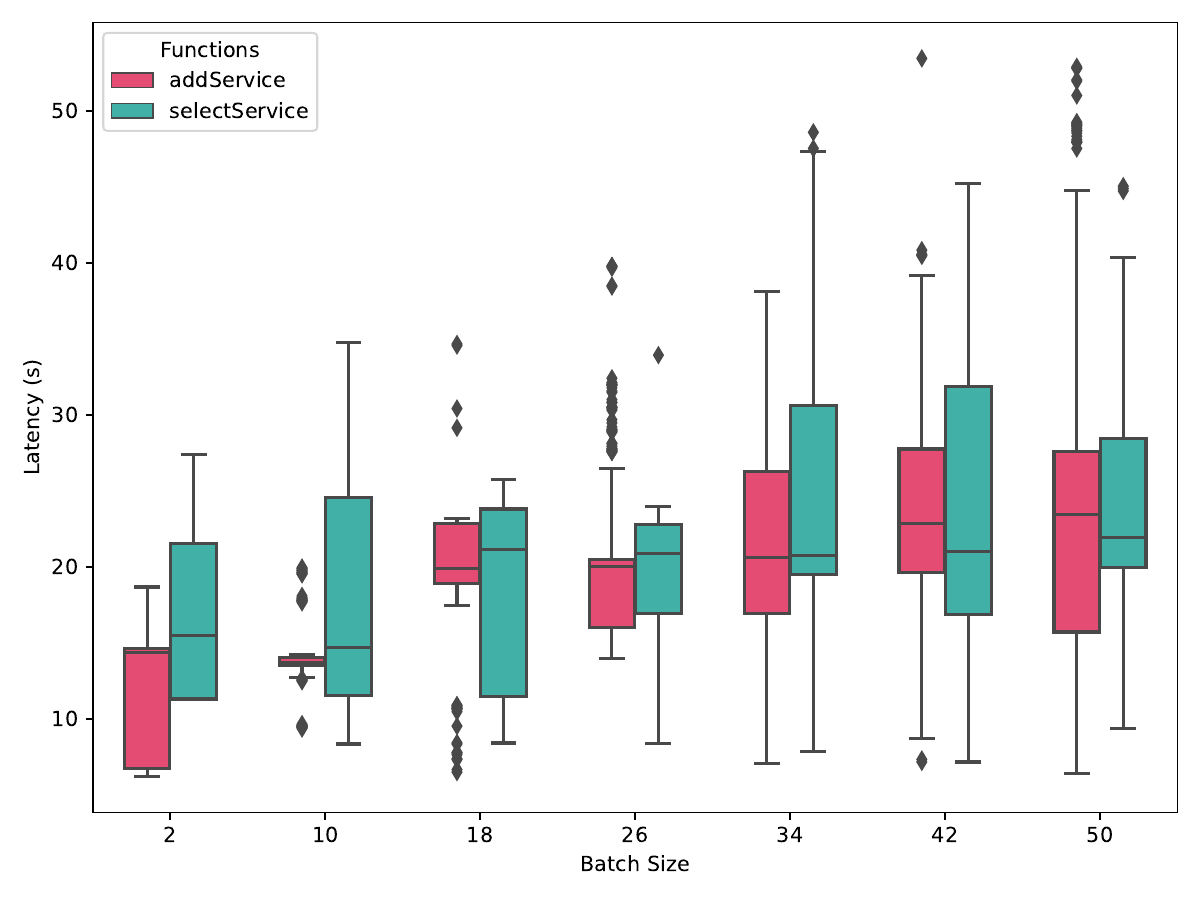}
    \caption{Transaction Latency for for \emph{Preliminary Agreement Phase} when Batch Size ranges from 2 to 50}
    \label{fig:latency-boxplot-phase1}
\end{figure}

As outlined in Section \ref{smart-contracts}, the \emph{Preliminary Agreement Phase} of the inter-provider DApp involves blockchain interactions during the addition and selection of services, incurring gas costs due to storage operations (Section \ref{smart-contracts-phase1}). Figure~\ref{fig:latency-boxplot-phase1} presents box plots illustrating transaction latencies (in seconds) when batch sizes range from 2 to 50. Overall, both \functionsc{addService} and \functionsc{selectService} exhibit a broad upward trend in latency as batch size increases, yet certain batches deviate from this pattern.

Table~\ref{table:addService-selectService} summarizes key parameters—block size (KB), gas price (Gwei), transaction count, and latency—for each tested batch size and function. Notably, even though larger batches (e.g., 34 or 42) often require processing more transactions, smaller batches do not always guarantee lower or more consistent latency. For example, Batch~2 sometimes has fewer transactions yet displays wider latency variability, indicating that transient network congestion or validator prioritization~\cite{eyal2014majority, decker2013information,sai2021performance} can overshadow the impact of batch size alone.

\begin{table*}[ht]
\centering
\caption{Summary of Quantitative Data by Batch Size for \functionsc{addService} and \functionsc{selectService} Functions}
\label{table:addService-selectService}
\begin{adjustbox}{width=1\textwidth, center}
\begin{tabular}{>{\raggedright\arraybackslash}p{2cm} | >{\centering\arraybackslash}p{1cm} >{\centering\arraybackslash}p{1.5cm} >{\centering\arraybackslash}p{1.5cm} | >{\centering\arraybackslash}p{1.5cm} >{\centering\arraybackslash}p{1.5cm} | >{\centering\arraybackslash}p{1.5cm} >{\centering\arraybackslash}p{1.5cm} | >{\centering\arraybackslash}p{1.5cm} >{\centering\arraybackslash}p{1.5cm}}
\toprule
\textbf{Function} & \textbf{Batch Size} & \multicolumn{2}{c|}{\textbf{Transaction Count}} & \multicolumn{2}{c|}{\textbf{Block Size (KB)}} & \multicolumn{2}{c|}{\textbf{Gas Price (Gwei)}} & \multicolumn{2}{c}{\textbf{Latency (s)}} \\
\cmidrule(lr){3-4} \cmidrule(lr){5-6} \cmidrule(lr){7-8} \cmidrule(lr){9-10}
 & & Mean & Std & Mean & Std & Mean & Std & Mean & Std \\
\midrule
\functionsc{addService}   & 2  & 127.9  & 17.17   & 181.26  & 62.17   & 15.59  & 1.74  & 12.51  & 4.22 \\
\functionsc{selectService} & 2  & 158.00 & 16.97   & 175.78  & 67.90   & 20.05  & 6.11  & 17.40  & 7.72 \\ \midrule
\functionsc{addService}   & 10 & 115.3  & 9.85    & 212.81  & 70.33   & 16.63  & 1.87  & 14.22  & 2.65 \\
\functionsc{selectService} & 10 & 119.06 & 32.52   & 142.33  & 67.84   & 22.89  & 5.06  & 17.59  & 8.25 \\ \midrule
\functionsc{addService}   & 18 & 142.63 & 32.00   & 148.65  & 45.20   & 14.16  & 0.60  & 19.93  & 4.47 \\
\functionsc{selectService} & 18 & 117.43 & 30.31   & 170.27  & 150.92  & 23.32  & 4.32  & 18.63  & 6.00 \\ \midrule
\functionsc{addService}   & 26 & 138.64 & 24.72   & 152.45  & 57.96   & 8.66   & 1.28  & 20.59  & 5.46 \\
\functionsc{selectService} & 26 & 124.69 & 27.51   & 112.06  & 35.07   & 23.67  & 3.58  & 19.27  & 4.87 \\ \midrule
\functionsc{addService}   & 34 & 132.77 & 25.25   & 154.34  & 68.31   & 5.35   & 0.41  & 21.51  & 5.99 \\
\functionsc{selectService} & 34 & 128.34 & 36.26   & 306.37  & 284.55  & 22.36  & 4.38  & 23.92  & 8.66 \\ \midrule
\functionsc{addService}   & 42 & 147.15 & 38.84   & 177.47  & 84.80   & 12.45  & 2.36  & 23.72  & 7.05 \\
\functionsc{selectService} & 42 & 123.06 & 28.25   & 159.33  & 119.53  & 22.69  & 4.26  & 23.06  & 8.13 \\ \midrule
\functionsc{addService}   & 50 & 120.13 & 34.75   & 171.62  & 71.26   & 8.03   & 0.99  & 23.46  & 9.62 \\
\functionsc{selectService} & 50 & 121.25 & 31.25   & 165.06  & 133.04  & 23.26  & 4.54  & 23.38  & 6.82 \\
\bottomrule
\end{tabular}
\end{adjustbox}
\end{table*}

For both functions, as shown in Figure~\ref{fig:latency-boxplot-phase1}, batches of 10, 18, and 26 stand out due to narrower Interquartile Ranges (IQR) (i.e., their box plots exhibit a smaller spread) and comparatively stable latencies, when compared to other batch sizes. By contrast, Batches 34 and 42 sometimes contain outliers with substantially higher latency, hinting at transient spikes in mempool congestion or fluctuations in validators’ block selection. Table~\ref{table:addService-selectService} corroborates these findings numerically. Interestingly, Batch~2 has a higher mean transaction count (especially for \functionsc{selectService}, at over 150) yet does not show proportionally lower latency, suggesting that fewer transactions alone cannot guarantee short or predictable turnaround times.

\functionsc{addService} starts with lower latencies (e.g., 12.51\,s at Batch~2) but climbs to around 23.46\,s by Batch~50. Meanwhile, \functionsc{selectService} begins near 17.40\,s at Batch~2 and reaches 23.38\,s at Batch~50 (Table~\ref{table:addService-selectService}). Despite this broad increase, stable gas prices (mostly 20–24\,gwei) often keep the median latency values in check, though occasional outliers underscore that network-level variability can override batch-size effects.

Although larger batches entail more data and can heighten resource demands \cite{sai2021performance,eyal2016bitcoinng}, smaller ones are not inherently optimal if sporadic congestion or validator behavior intervenes~\cite{eyal2014majority, decker2013information}. Real-time conditions—such as block size usage, gas price stability, and the overall transaction count—can produce sudden spikes or dips. This partially explains why Batches~10 and~18 remain relatively stable (i.e., they exhibit a lower standard deviation and narrower IQR in Table~\ref{table:addService-selectService}), whereas Batches~2, 34, and~42 show greater variability (their higher standard deviations indicate latencies are more widely dispersed). However, to fully parse these trends and confirm which parameters most strongly affect latency, we conduct a deeper statistical analysis in the following section. There, we categorize transactions by gas price, block size, and transaction count quintiles, applying omnibus tests to quantify their impacts on latency distributions.

\begin{table*}[ht]
\centering
\caption{Summary of Quantitative Data by Batch Size for \functionsc{registerBreach} and \functionsc{calculatePenalty}}
\label{table:penalty}
\begin{adjustbox}{width=1\textwidth, center}
\begin{tabular}{>{\raggedright\arraybackslash}p{2.5cm} | >{\centering\arraybackslash}p{1cm} >{\centering\arraybackslash}p{1.5cm} >{\centering\arraybackslash}p{1.5cm} | >{\centering\arraybackslash}p{1.5cm} >{\centering\arraybackslash}p{1.5cm} | >{\centering\arraybackslash}p{1.5cm} >{\centering\arraybackslash}p{1.5cm} | >{\centering\arraybackslash}p{1.5cm} >{\centering\arraybackslash}p{1.5cm}}
\toprule
\textbf{Function} & \textbf{Batch Size} & \multicolumn{2}{c|}{\textbf{Transaction Count}} & \multicolumn{2}{c|}{\textbf{Block Size (KB)}} & \multicolumn{2}{c|}{\textbf{Gas Price (Gwei)}} & \multicolumn{2}{c}{\textbf{Latency (s)}} \\
\cmidrule(lr){3-4} \cmidrule(lr){5-6} \cmidrule(lr){7-8} \cmidrule(lr){9-10}
 & & Mean & Std & Mean & Std & Mean & Std & Mean & Std \\
\midrule
\functionsc{registerBreach} & 2 & 94.07 & 22.05 & 219.59 & 172.93 & 1.50 & 0.00 & 12.74815 & 4.39672 \\ 
\functionsc{calculatePenalty} & 2 & 90.50 & 15.00 & 227.68 & 210.60 & 1.50 & 0.00 & 10.99975 & 1.71643 \\ \midrule
\functionsc{registerBreach} & 10 & 113.23 & 27.41 & 243.16 & 145.19 & 1.50 & 0.00 & 15.96191 & 8.59601 \\ 
\functionsc{calculatePenalty} & 10 & 103.95 & 18.21 & 257.06 & 214.64 & 1.50 & 0.00 & 12.09535 & 4.42006 \\ \midrule
\functionsc{registerBreach} & 18 & 120.01 & 22.28 & 219.54 & 169.38 & 21.16 & 4.14 & 17.44435 & 9.10133 \\ 
\functionsc{calculatePenalty} & 18 & 114.78 & 22.60 & 210.28 & 182.63 & 20.77 & 4.08 & 13.30501 & 10.20327 \\ \midrule
\functionsc{registerBreach} & 26 & 140.80 & 48.57 & 220.86 & 146.88 & 31.73 & 13.19 & 21.24473 & 11.18447 \\ 
\functionsc{calculatePenalty} & 26 & 150.30 & 46.66 & 217.45 & 126.48 & 33.12 & 14.02 & 18.59561 & 6.92867 \\ \midrule
\functionsc{registerBreach} & 34 & 115.04 & 27.61 & 230.63 & 149.41 & 33.41 & 7.32 & 21.15253 & 10.38157 \\ 
\functionsc{calculatePenalty} & 34 & 117.63 & 36.76 & 247.94 & 191.65 & 31.50 & 4.91 & 16.53323 & 10.68454 \\ \midrule
\functionsc{registerBreach} & 42 & 131.61 & 41.48 & 230.53 & 162.30 & 53.55 & 28.10 & 23.51004 & 14.75455 \\ 
\functionsc{calculatePenalty} & 42 & 119.55 & 29.30 & 239.88 & 183.28 & 51.49 & 29.44 & 15.57647 & 7.73852 \\ \midrule
\functionsc{registerBreach} & 50 & 121.74 & 39.24 & 265.95 & 218.36 & 10.15 & 6.76 & 24.69966 & 14.85866 \\ 
\functionsc{calculatePenalty} & 50 & 113.33 & 35.48 & 233.65 & 175.64 & 10.08 & 6.84 & 17.34491 & 11.12110 \\
\bottomrule
\end{tabular}
\end{adjustbox}
\end{table*}
In particular, \functionsc{selectService} latency tracks the time from the consumer’s function call—where \(\texttt{\$S\_ID}=5\) is an example of the requested service identifier—to the on-chain finalization. Maintaining moderate gas prices helps reduce competition among transactions, but sporadic outliers (for instance, in Batch~42 with a mean latency of 23.92\,s) reveal that even small network fluctuations can disproportionately delay certain transactions. By contrast, \functionsc{addService} exhibits a similar pattern—some lower-latency batches intermixed with outliers at higher batch sizes—highlighting that both storage-heavy (adding services) and selection-based operations are subject to these network dynamics.

In summary, while latency generally increases with larger batches, real-time network factors can override any simple “batch size” rule. The subsequent statistical section delves into these factors—gas price, block size, transaction count—to validate whether the observed anomalies have a significant effect on transaction latency.

\begin{figure}[t!]
    \centering
    \includegraphics[width=1.1\linewidth]{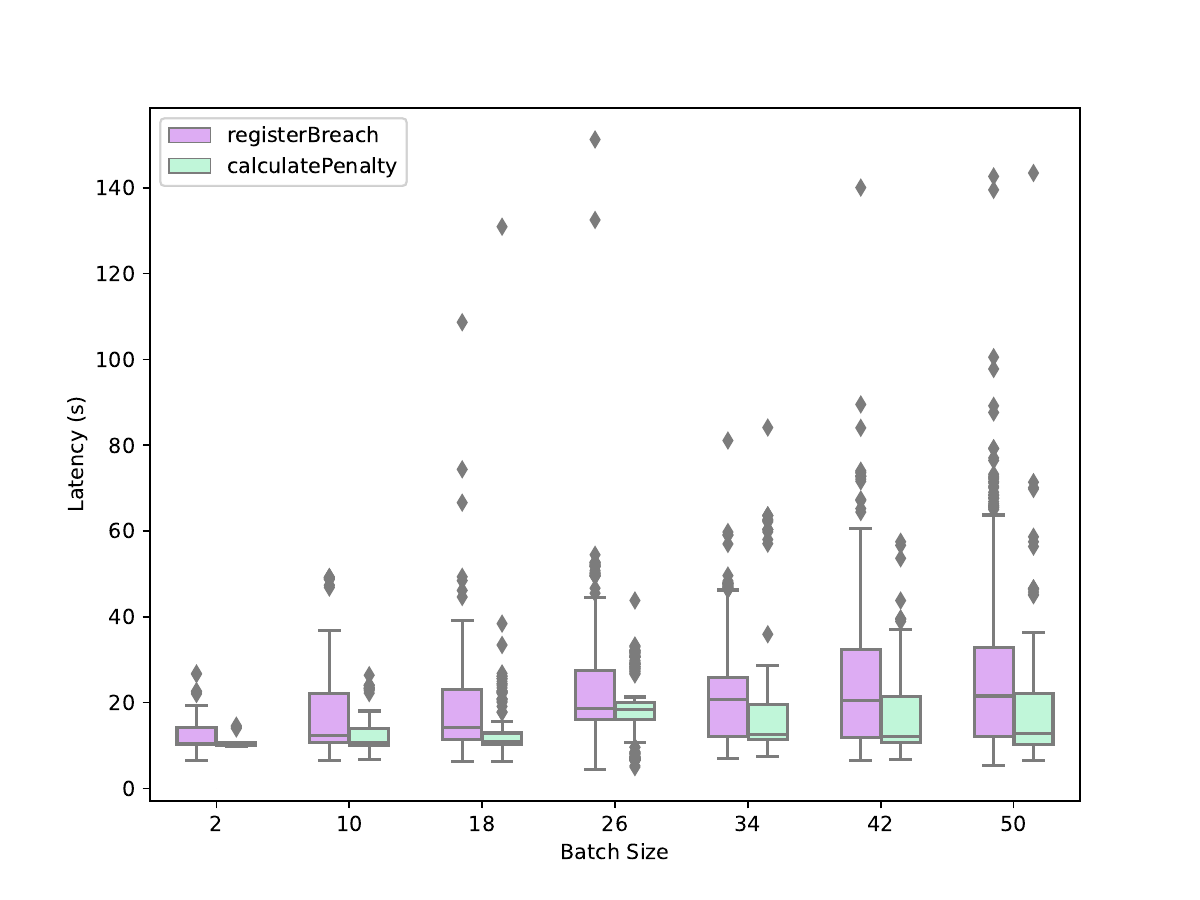}
    \caption{Transaction Latency for \emph{Enforcement Phase} when Batch Size ranges from 2 to 50}
    \label{fig:latency-boxplot-phase2}
\end{figure}

%\subsubsection{Performance Analysis of \texttt{addService} and \texttt{selectService}}

\subsubsection{\functionsc{registerBreach} and \functionsc{calculatePenalty}}

%\JM{this section sounds a bit repetitive with the previous one. I think that if we decide to keep it, we should emphasize the new information it brings compared to the previous one. Did you compare? Do we draw exactly the same conclusions and observe the same behavior in the same batches? We should also explain why these two subsections and why you grouped the smart contracts like this. Is there a reason for grouping like this? Then, explain. But the main point is: I would focus on the new additional info that this analysis provides compared to the previous section and say that for the rest, the conclusions are the same.}

Compared to the \emph{Preliminary Agreement Phase} (\functionsc{addService}, \functionsc{selectService}), where each transaction tends to involve higher gas usage per call (e.g., initializing provider services or selecting a new service for the first time) but occurs more sporadically, the \emph{Enforcement Phase} typically uses lower-gas calls that can be triggered repeatedly in quick succession. This design difference underpins a distinct latency profile.

As outlined in Section \ref{smart-contracts}, the \emph{Enforcement Phase} of the inter-provider DApp involves blockchain interactions during the registration of breaches and calculation of penalties, incurring gas costs due to storage operations (Section \ref{smart-contracts-phase2}). 
The \functionsc{registerBreach} function in the \contract{RegisterBreach.sol} smart contract increments a provider’s breach count each time a violation is detected, ultimately triggering \functionsc{calculatePenalty} once the threshold of \(B_{\text{max}}=3\) is reached. Figure~\ref{fig:latency-boxplot-phase2} displays box plots of the measured latencies (in seconds) for these functions when batch sizes range from 2 to 50. Although an upward trend in median latency emerges with increasing batch size, certain batches deviate from this pattern, pointing to factors beyond simple “more transactions = higher latency.”

Table~\ref{table:penalty} provides supporting quantitative details—transaction count, block size, gas price, and latency—for each batch. At smaller batch sizes (e.g., Batch~2), \functionsc{registerBreach} averages around 12.74\,s, while \functionsc{calculatePenalty} hovers near 11\,s, both with relatively modest variability. One might expect fewer transactions to result in uniformly low latencies, yet these smaller batches occasionally show outlier points above 20\,s, implying that minor periods of mempool congestion or validator reordering~\cite{eyal2014majority, decker2013information} can overshadow the simplicity of a small batch. 

Moving to Batch~10 and Batch~18, latencies generally stay in the 12--17\,s range for \functionsc{registerBreach} and slightly lower for \functionsc{calculatePenalty}, though each function produces sporadic outliers in certain trials. By contrast, Batches~26, 34, and 42 often exhibit a distinct cluster of higher-latency transactions—some reaching beyond 40 or even 100\,s—indicating that periodic spikes in network activity (e.g., concurrent transactions from other users) can disproportionately affect certain batches. These larger batches also feature more repeated calls to \functionsc{registerBreach} or \functionsc{calculatePenalty}, which may temporarily push block resources close to their limits.

In contrast, \functionsc{addService} and \functionsc{selectService} usually involve one-time additions or selections carrying a higher per-transaction gas cost (Section~\ref{smart-contracts-phase1}). By comparison, \functionsc{registerBreach} often uses less gas per call but can happen multiple times for the same provider. When many breaches converge, \functionsc{calculatePenalty} is triggered repeatedly, potentially causing latency spikes unrelated to the total batch size alone.

Interestingly, Batch~50 does not always yield the worst latencies despite being the largest group; \functionsc{registerBreach} at Batch~50 averages about 24.70\,s, whereas some smaller batches (e.g., Batch~42) record similar or even higher average times. This observation reinforces that real-time network conditions—and not merely the raw count of function calls—can tip the latency scales one way or another. Gas price and block size variations, summarized in Table~\ref{table:penalty}, appear to influence these outcomes in subtle ways, but their exact impact requires further breakdown. Overall, \functionsc{registerBreach} and \functionsc{calculatePenalty} follow a broad “more calls, more latency” progression yet reveal exceptions at both small and large batch sizes. A closer look at concurrent factors—such as gas price fluctuations, block occupancy, and network congestion—will clarify why these anomalies occur. As with the \emph{Preliminary Agreement Phase}, the subsequent statistical analysis section addresses these finer points by categorizing transactions into relevant quintiles (for gas price, block size, and transaction count), then quantitatively evaluating their relationship to latency.

Finally, the threshold-based nature of \functionsc{registerBreach} and \functionsc{calculatePenalty} introduces outlier scenarios absent in \functionsc{addService} or \functionsc{selectService}. Once \(\texttt{maxBreach}=3\) is reached for multiple providers in close succession, \texttt{calculatePenalty} can be invoked repeatedly within a short window, amplifying latency beyond what higher-gas but one-off operations in the \emph{Preliminary Agreement Phase} experienced. This dynamic justifies analyzing the \emph{Enforcement Phase} separately, as it reveals a distinct mechanism driving extreme delays that cannot be captured by batch size considerations alone.

\begin{table*}[ht]
\centering
\caption{Kruskal-Wallis H-test Results for Preliminary Agreement and \emph{Enforcement Phase}}
\label{table:combined_kruskal_wallis}
\begin{adjustbox}{max width=\textwidth}
\begin{tabular}{p{2.1cm}|p{1.3cm}p{1cm}p{2.3cm}|p{1.3cm}p{1cm}p{2.3cm}}
\toprule
\textbf{Feature} & \multicolumn{3}{c|}{\textbf{Preliminary Agreement Phase}} & \multicolumn{3}{c}{\textbf{Enforcement Phase}} \\ \cmidrule(lr){2-4} \cmidrule(lr){5-7}
 & \textbf{H-statistic} & \textbf{p-value} & \textbf{Interpretation} & \textbf{H-statistic} & \textbf{p-value} & \textbf{Interpretation} \\ \midrule
Gas Price (Gwei) & 31.8553 & 0.000002 & \textbf{Significant} differences & 144.47 & 0.0000 & \textbf{Significant} differences \\
Block Size (KB) & 107.8159 & 0.0 & \textbf{Significant} differences & 10.56 & 0.0320 & \textbf{Significant} differences \\
Transaction Count & 211.2898 & 0.0 & \textbf{Significant} differences & 21.90 & 0.0002 & \textbf{Significant} differences \\
\bottomrule
\end{tabular}
\end{adjustbox}
\end{table*}

\subsection{Dependency of Transaction Latency on Blockchain Parameters}

\subsubsection{Gas Price}
As explained earlier in Section \ref{parameters} gas price measures the per-unit cost a user offers to pay for the computational resources consumed by their transaction. %Consequently, higher gas prices generally lead to faster confirmations (lower latency), whereas lower gas prices risk longer wait times.

The results in Table~\ref{table:combined_kruskal_wallis} reveal that gas price carries varying levels of significance between the two phases. In the \emph{Preliminary Agreement Phase}, its Kruskal–Wallis H-statistic is 31.8553, which is lower than those of block size (107.8159) and transaction count (211.2898). By contrast, in the \emph{Enforcement Phase}, gas price emerges with the highest H-statistic of 144.47, surpassing transaction count (21.90) and block size (10.56). The combined post-hoc tables (Tables~\ref{table:combined_post_hoc_phase1} and \ref{table:combined_posthoc_analysis_for_phase2} corroborate these findings, showing that gas price produces different latency distributions (all with $p < 0.05$) but with differing effect sizes in each phase.

\begin{figure*}[]
    \centering
    \includegraphics[width=1\linewidth]{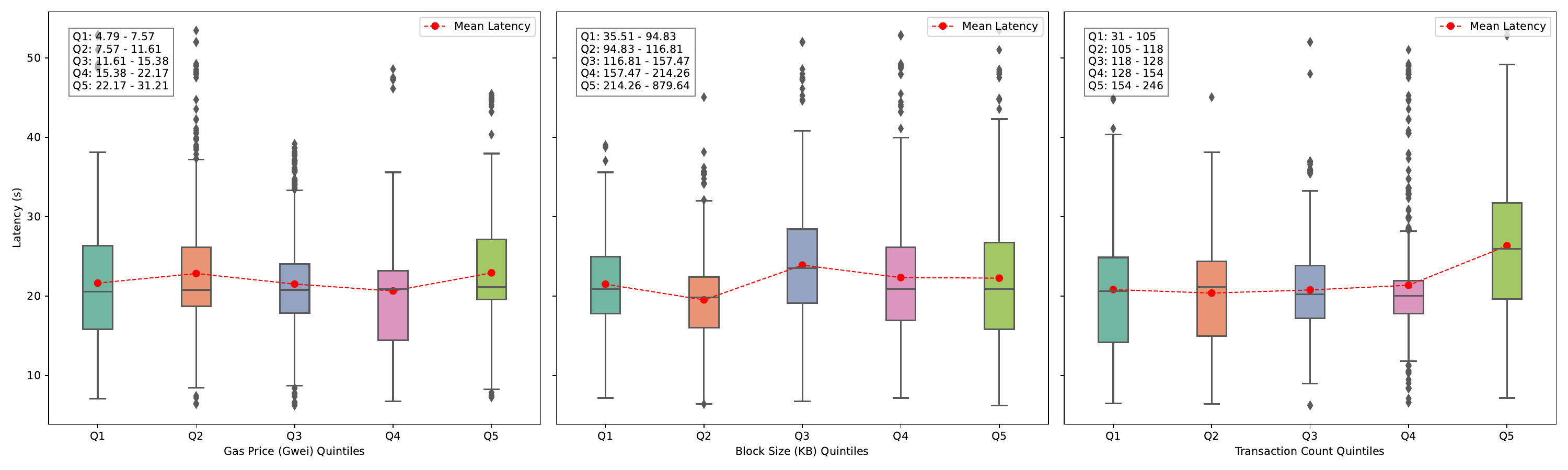}
    \caption{Box Plots of Observed Latency (s) Across Quintiles for Gas Price (Gwei), Block Size (KB), and Transaction Count in the \textbf{\emph{Preliminary Agreement Phase}}. Each panel displays latency against quintiles of each parameter, showing the range of values within each quintile: Gas Price (left panel), Block Size (middle panel), and Transaction Count (right panel). The red dashed line represents the mean latency across quintiles}
    \label{fig:phase1-latency}
\end{figure*}

For a visual comparison of these quintile-based differences in the \emph{Preliminary Agreement Phase}, see Figure~\ref{fig:phase1-latency}. 
The left panel illustrates how Gas Price quintiles (Q1--Q5) influence latency outcomes, aligning with the small-to-medium effect sizes reported in Table~\ref{table:combined_post_hoc_phase1}.

In the Gas Price panel of Figure~\ref{fig:phase1-latency} (left), Q4 shows the lowest median latency, while Q2 and Q5 appear slightly higher. 
Table~\ref{table:combined_post_hoc_phase1} confirms these observations: for instance, Q2 vs.\ Q4 is significant (p=0.0002, Cliff's Delta=0.14), indicating Q4’s latency distribution is modestly lower than Q2’s. 
Similarly, Q4 vs.\ Q5 (p=0.0000, $\Delta=-0.17$) highlights that even very high gas bids (Q5) do not necessarily deliver the fastest confirmations.

\begin{table*}[t!]
\centering
\caption{Dunn’s Test Post-hoc Analysis Comparing Latency Across Quintiles for Gas Price, Block Size, and Transaction Count During the \textbf{\emph{Preliminary Agreement Phase}}. Each cell reports whether the difference is statistically significant (Yes/No), the p-value of the test, and the effect size (Cliff’s Delta), categorizing the magnitude of difference between quintiles}
\label{table:combined_post_hoc_phase1}
\begin{adjustbox}{max width=\textwidth}
\begin{tabular}{p{1.8cm}|p{0.8cm}p{2cm}p{2.5cm}|p{0.8cm}p{2cm}p{2.5cm}|p{0.8cm}p{2cm}p{2.5cm}}
\toprule
\textbf{Comparison} & \multicolumn{3}{c|}{\textbf{Gas Price (Gwei)}} & \multicolumn{3}{c|}{\textbf{Block Size (KB)}} & \multicolumn{3}{c}{\textbf{Transaction Count}} \\ 
\cmidrule(lr){2-4} \cmidrule(lr){5-7} \cmidrule(lr){8-10}
 & \textbf{Yes/No} & \textbf{p-value} & \textbf{Effect size (Delta)} 
 & \textbf{Yes/No} & \textbf{p-value} & \textbf{Effect size (Delta)} 
 & \textbf{Yes/No} & \textbf{p-value} & \textbf{Effect size (Delta)} \\ 
\midrule
Q1 vs Q2 & No  & 0.5639 & Not significant 
         & Yes & 0.0   & 0.2 (Small) 
         & No  & 1.0   & Not significant \\
Q1 vs Q3 & No  & 1.0000 & Not significant 
         & Yes & 0.0   & -0.18 (Small)   % corrected from Negligible to Small
         & No  & 1.0   & Not significant \\
Q1 vs Q4 & No  & 0.1851 & Not significant 
         & No  & 1.0   & Not significant 
         & No  & 1.0   & Not significant \\
Q1 vs Q5 & No  & 0.0711 & Not significant 
         & No  & 1.0   & Not significant 
         & Yes & 0.0   & -0.37 (Medium)  \\
Q2 vs Q3 & No  & 0.1651 & Not significant 
         & Yes & 0.0   & -0.37 (Medium) 
         & No  & 1.0   & Not significant \\
Q2 vs Q4 & Yes & 0.0002 & 0.14 (Negligible) 
         & Yes & 0.0   & -0.2 (Small) 
         & No  & 1.0   & Not significant \\
Q2 vs Q5 & No  & 1.0000 & Not significant 
         & Yes & 0.0   & -0.17 (Small)   % corrected from Negligible to Small
         & Yes & 0.0   & -0.43 (Medium)  \\
Q3 vs Q4 & No  & 0.6889 & Not significant 
         & Yes & 0.0   & 0.16 (Small)    % corrected from Negligible to Small
         & No  & 1.0   & Not significant \\
Q3 vs Q5 & Yes & 0.0152 & -0.12 (Negligible) 
         & Yes & 0.0   & 0.15 (Small)    % corrected from Negligible to Small
         & Yes & 0.0   & -0.41 (Medium)  \\
Q4 vs Q5 & Yes & 0.0000 & -0.17 (Small)   % corrected from Negligible to Small
         & No  & 1.0   & Not significant 
         & Yes & 0.0   & -0.39 (Medium)  \\ 
\bottomrule
\end{tabular}
\end{adjustbox}
\end{table*}

In the \emph{Preliminary Agreement Phase}, \functionsc{addService} and \functionsc{selectService} operations each consume roughly 140k–160k gas (Table~\ref{tab:smart_contracts} in earlier sections). These storage-heavy transactions can saturate block resources and inflate latencies in larger batches~\cite{croman2016scaling, decker2013information,luu2015incentives}. Consequently, even if certain users bid moderately higher or lower gas prices, the net effect on latency is overshadowed by the greater impact of block capacity and transaction volume. Gas price differences remain statistically significant according to the Kruskal–Wallis test, yet the lower H-statistic indicates a smaller association with latency outcomes than block size or transaction count in this phase. %\JM{if you say this, it means that you have a way of measuring the individual contribution of components of variance. If this is not the case, I would rephrase.} 
The Hardhat framework, connected via Alchemy to Sepolia, automatically sets gas prices in line with current network conditions; users do not manually override fees, so the observed gas price fluctuations stem from real-time base-fee and priority-tip adjustments~\cite{buterin2019eip1559, eyal2014majority}.

\begin{figure*}
    \centering
    \includegraphics[width=1\linewidth]{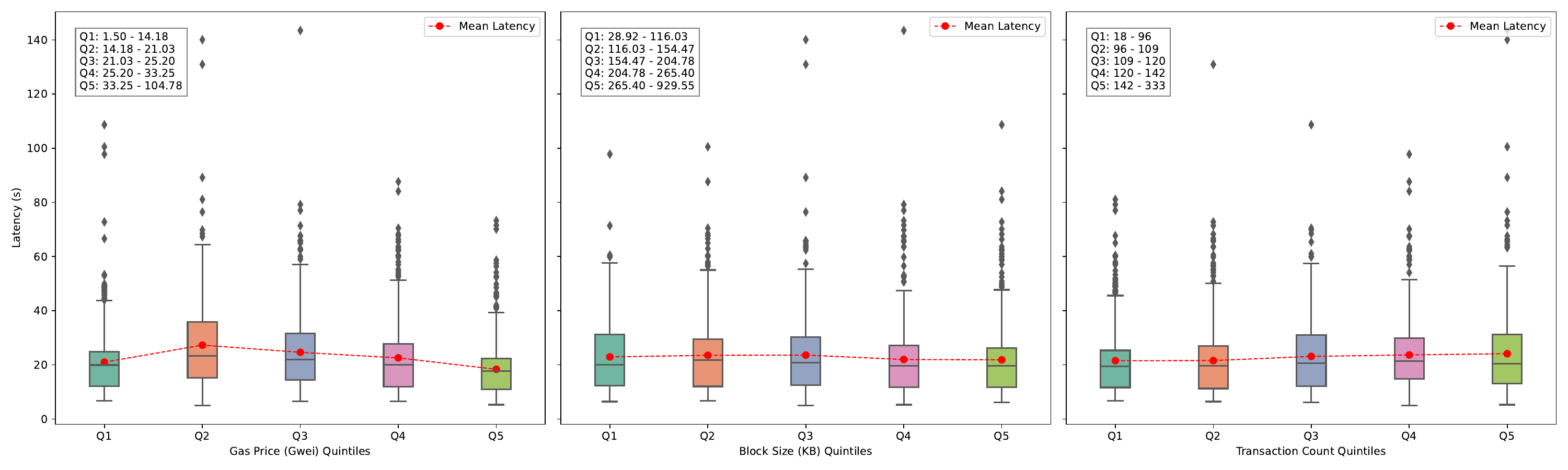}
    \caption{Box Plots of Observed Latency (s) Across Quintiles for Gas Price (Gwei), Block Size (KB), and Transaction Count in the \textbf{\textit{Enforcement Phase}}. Each panel displays latency against quintiles of each parameter, showing the range of values within each quintile: Gas Price (left panel), Block Size (middle panel), and Transaction Count (right panel). The red dashed line represents the mean latency across quintiles}
    \label{fig:low-gas-latency}
\end{figure*}

% Combined Table: Post-hoc Analysis for Gas Price, Block Size, and Transaction Count
\begin{table*}[t]
\centering
\caption{Dunn’s Test Post-hoc Analysis Comparing Latency Across Quintiles for Gas Price, Block Size, and Transaction Count During the \textbf{\textit{Enforcement Phase}}. Each cell reports whether the difference is statistically significant (Yes/No), the p-value of the test, and the effect size (Cliff’s Delta), categorizing the magnitude of difference between quintiles}
\label{table:combined_posthoc_analysis_for_phase2}
\begin{adjustbox}{max width=\textwidth}
\begin{tabular}{p{1.8cm}|p{1.5cm}p{1.9cm}p{2.2cm}|p{1.5cm}p{1.9cm}p{2.2cm}|p{1.5cm}p{1.9cm}p{2.2cm}} 
\toprule
\textbf{Comparison} 
& \multicolumn{3}{c|}{\textbf{Gas Price (Gwei)}} 
& \multicolumn{3}{c|}{\textbf{Block Size (KB)}} 
& \multicolumn{3}{c}{\textbf{Transaction Count}} 
\\ 
\cmidrule(lr){2-4} \cmidrule(lr){5-7} \cmidrule(lr){8-10}
 & \textbf{Yes/No} & \textbf{p-value} & \textbf{Cliff's Delta} 
 & \textbf{Yes/No} & \textbf{p-value} & \textbf{Cliff's Delta} 
 & \textbf{Yes/No} & \textbf{p-value} & \textbf{Cliff's Delta} \\ 
\midrule
Q1 vs Q2 & Yes & 0.0     & -0.236 (Small) 
         & No  & 1.0     & -0.028 (Negligible) 
         & No  & 1.0     &  0.013 (Negligible) \\
Q1 vs Q3 & Yes & 0.000004 & -0.165 (Small)  % corrected from Negligible to Small
         & No  & 1.0     & -0.025 (Negligible) 
         & No  & 0.22443 & -0.074 (Negligible) \\
Q1 vs Q4 & No  & 1.0      & -0.054 (Negligible) 
         & No  & 1.0      &  0.042 (Negligible) 
         & Yes & 0.01307  & -0.117 (Negligible)\\
Q1 vs Q5 & Yes & 0.00017  &  0.165 (Small)  % corrected from Negligible to Small
         & No  & 1.0      &  0.054 (Negligible) 
         & Yes & 0.02662  & -0.106 (Negligible)\\
Q2 vs Q3 & No  & 0.35314  &  0.086 (Negligible) 
         & No  & 1.0      &  0.002 (Negligible) 
         & No  & 0.117    & -0.085 (Negligible)\\
Q2 vs Q4 & Yes & 0.0      &  0.187 (Small) 
         & No  & 0.34593  &  0.073 (Negligible) 
         & Yes & 0.00589  & -0.118 (Negligible)\\
Q2 vs Q5 & Yes & 0.0      &  0.357 (Medium) 
         & No  & 0.12634  &  0.088 (Negligible) 
         & Yes & 0.01231  & -0.111 (Negligible)\\
Q3 vs Q4 & Yes & 0.01359  &  0.113 (Negligible) 
         & No  & 0.41997  &  0.072 (Negligible) 
         & No  & 1.0      & -0.032 (Negligible)\\
Q3 vs Q5 & Yes & 0.0      &  0.313 (Small)   % corrected from Medium to Small
         & No  & 0.15805  &  0.083 (Negligible) 
         & No  & 1.0      & -0.026 (Negligible)\\
Q4 vs Q5 & Yes & 0.0      &  0.198 (Small) 
         & No  & 1.0      &  0.014 (Negligible) 
         & No  & 1.0      &  0.010 (Negligible)\\ 
\bottomrule
\end{tabular}
\end{adjustbox}
\end{table*}

By contrast, in the \emph{Enforcement Phase}, gas price becomes the dominant factor, reflected by its highest H-statistic (144.47). Here, \functionsc{registerBreach} and \functionsc{calculatePenalty} use significantly lower per-call gas ($\approx 44,058$–$49,143$), though \functionsc{registerBreach} is invoked three times per breach scenario. These smaller, repeated transactions cause total block usage to remain within manageable limits, thus giving gas price a clearer role in determining which transactions are prioritized. Hardhat’s network-driven fee mechanism thus leads to wider variability in how quickly each breach or penalty call confirms, illustrating that when blocks are less heavily saturated, fee bidding plays a decisive role in mempool inclusion and eventual block proposal~\cite{gervais2016security}.

Figure~\ref{fig:low-gas-latency} illustrates how Gas Price 
(left panel) strongly influences latency in the \emph{Enforcement Phase}. Notably, Q5—the highest gas tier—shows the lowest median latency, aligned with the post-hoc comparisons in 
Table~\ref{table:combined_posthoc_analysis_for_phase2}.

The post-hoc analysis for the \emph{Enforcement Phase} (Table~\ref{table:combined_posthoc_analysis_for_phase2}) indicates that moderate gas-price quintiles (e.g., Q2 or Q3) can yield significantly lower or comparable latencies than the highest tier (Q5). For instance, comparing Q2 vs.\ Q5 yields a p-value of 0.0 with a medium effect size (Cliff’s Delta = 0.357), suggesting that simply paying the extreme gas-price tier does not always result in faster confirmations. One explanation is that short-lived (ephemeral) surges in the mempool can negate the advantage of overbidding if multiple high-fee transactions arrive simultaneously. In addition, validator scheduling may spread out transactions across blocks in ways that dilute the expected speed gain from overpaying~\cite{buterin2019eip1559}. Thus, while gas price remains a dominant factor in this phase, extreme bids do not guarantee minimal latency in every scenario.

\subsubsection{Block Size}
Block size measures the total data capacity allocated to a single block (e.g., in bytes). It represents how much transaction data and metadata can fit into each mined block as explained earlier in section \ref{parameters}.

In the \emph{Preliminary Agreement Phase}, block size exhibits a high H-statistic of 107.8159, second only to transaction count. Figure~\ref{fig:phase1-latency} (middle panel) shows that blocks around Q2 (roughly 95–116~KB) consistently yield lower mean latencies, whereas blocks in Q3 or above can reach 150–200+~KB and produce higher average confirmation times. Post-hoc comparisons highlight significant differences in latency distributions once block size surpasses a certain threshold, suggesting that network propagation delays and validation overhead become pivotal~\cite{croman2016scaling, decker2013information}. Because \functionsc{addService}’s storage-heavy operations frequently consume over 80\% of a block’s data capacity (Section~\ref{smart-contracts-phase1}), the block size becomes a bottleneck by delaying network propagation and validation. Larger blocks (e.g., 150–200+ KB) require more time to propagate across nodes and validate, inflating latency even when gas prices or transaction counts are stable.

However, the \emph{Enforcement Phase} reveals a far smaller H-statistic for block size (10.56), indicating that while it remains significant ($p = 0.0320$) [see Table \ref{table:combined_kruskal_wallis}], it is comparatively less influential than gas price or transaction count in explaining latency variance. Because the repeated yet lower-gas calls (\functionsc{registerBreach} and \functionsc{calculatePenalty}) typically keep block sizes below 150--200~KB, Phase~2 rarely hits the extremes that degrade network propagation speeds. As Figure~\ref{fig:low-gas-latency} (middle panel) suggests, most block-size quintiles stay within a range (For example, the lower quintiles of Phase 2 span 28.92–116.63~KB (Q1) and 116.63–154.47~KB (Q2), meaning that a large proportion of blocks do not approach the higher sizes that are known to cause network propagation and validation delays) that does not markedly inflate latency, so although block size differentials are measurable, they do not cause the pronounced delays observed in Phase~1. Consequently, although differences in block size are measurable, they do not lead to the pronounced latency delays seen in Phase 1, where blocks often exceed 150–200~KB.
%\JM{do you mean that the block size in this case have moderate size and, as a consequence, latency shows less variance? If this is what you want to say, please rephrase. If this is not the case, then I do not see moderate block size in this figure.} 
Consequently, block size differentials, although measurable, do not create as pronounced a delay in Phase 2.

This contrast between the two phases underscores that block size is context-dependent: heavy contract operations like adding and selecting services can balloon a block’s raw data footprint, making block size a key latency determinant, whereas lighter calls—even repeated—leave more slack for the network to handle those blocks efficiently \cite{chen2017contracts,sai2021performance}.

\subsubsection{Transaction Count}
As explained in Section \ref{parameters}, transaction count measures the number of individual transactions included within a block. It reflects how many discrete operations are recorded in that block.

Transaction count exhibits the largest H-statistic (211.2898) in the \emph{Preliminary Agreement Phase} and a more modest yet still significant value (21.90) in the \emph{Enforcement Phase}. This outcome aligns directly with the nature of the transactions. In Phase 1, \functionsc{addService} and \functionsc{selectService} each require higher gas and often appear in larger batch sizes, leading to blocks that combine many big transactions. As soon as the number of concurrent transactions rises (e.g., 150–200 in one block range), latencies show a pronounced jump~\cite{eyal2014majority}. The post-hoc table for Phase 1 (Table~\ref{table:combined_post_hoc_phase1}) confirms that Q5 (154–246 transactions) correlates with medium effect-size differences in latency (e.g., $\Delta \approx -0.39$ to $-0.43$ vs.\ lower quintiles).

In Phase 2, transaction count remains significant but loses its top position to gas price. Since each function call is relatively lighter ($\approx 44k$–$49k$ gas), even if many transactions stack in a block, the overall block size rarely approaches the critical. By critical levels, we refer to the network capacity thresholds where block size becomes so large that it significantly degrades performance (e.g., causing increased propagation and validation delays). In Phase 1, when blocks reach around 150–200 transactions or larger sizes, these thresholds are met, leading to pronounced latency spikes. levels seen in Phase 1. As a result, competition among transactions in the mempool centers more on how much gas fee each is paying rather than on whether a block can physically accommodate them. The boxplot for Phase 2 (Table~\ref{table:combined_posthoc_analysis_for_phase2}) still shows a jump in latency at higher transaction counts, but it is overshadowed by the more substantial effect of gas price bidding~\cite{gervais2016security, decker2013information}. Only in the densest scenario (Q5 with 142–333 transactions) do we see a notable rise in median latencies, reflecting the upper boundary of the network’s throughput capacity in the \emph{Enforcement Phase}.

Ultimately, transaction count in each phase demonstrates consistent patterns: significant differences in latency distributions when blocks are overfilled. The degree of that effect, however, depends on how large or small each call is. \emph{Preliminary Agreement Phase} amplifies transaction-count effects due to heavier calls, whereas the \emph{Enforcement Phase}—though repeated—spreads lighter calls over multiple blocks, reducing the direct impact of concurrency peaks.

\subsection{Discussion and Lesson Learned}
The empirical results reveal that gas price, block size, and transaction count influence on-chain latencies, but these parameters assert themselves differently depending on the contract complexity (Tables~\ref{table:combined_post_hoc_phase1} and \ref{table:combined_posthoc_analysis_for_phase2}). For example, as presented in this use case, for \emph{Preliminary Agreement Phase} (i.e., \functionsc{addService}, \functionsc{selectService}), larger storage-heavy transactions—roughly 140k–160k gas each—tend to saturate the block’s capacity. This makes block size and transaction count the most decisive latency drivers, producing medium effect sizes in the range of $-0.37$ to $-0.43$ and overshadowing any advantage gained by paying higher fees. Conversely, in the \emph{Enforcement Phase} (i.e., \functionsc{registerBreach}, \functionsc{calculatePenalty}), these calls are lighter, seldom saturating block limits, which leads to gas price emerging as the strongest factor; here, higher bids often correlate with faster confirmations (up to $\Delta \approx 0.36$), validating the standard intuition that “paying more yields faster inclusion” \cite{buterin2019eip1559, gervais2016security, monrat2019survey}.

Although the statistical significance is apparent in many comparisons (p-values near 0.0 for Q1 vs. Q5 contrasts), the practical impact is often only moderate. For instance, certain Q3 vs. Q4 or Q4 vs. Q5 comparisons in Table~\ref{table:combined_post_hoc_phase1} show no significant difference (p-values near 1.0), highlighting that large jumps in block size or transaction count do not \emph{always} guarantee higher latency in a uniform manner. The box plots in Figures~\ref{fig:phase1-latency} and \ref{fig:low-gas-latency} reinforce these observations by showing overlapping latencies: some top-quintile gas-price transactions (Q5) can lag behind lower-quintile ones (Q3, Q4) when transient concurrency spikes or validator scheduling nuances occur. Such observations underscore that purely theoretical heuristics—like “whoever pays more always wins” or “larger blocks always confirm last”—are inadequate to predict actual on-chain timing, echoing well-documented effects of concurrency, node connectivity, and proof-of-stake scheduling \cite{eyal2014majority, decker2013information, chen2017contracts, croman2016scaling}.

Taken together, the data reveal that \emph{each parameter can become dominant} under certain conditions, aligning with the broader consensus in blockchain research that real-world performance emerges from complex interactions among concurrency, transaction complexity, and protocol-level rules \cite{wood2014ethereum, buterin2014ethereum}. In moments when multiple heavy transactions cluster in the same block (as in the \emph{Preliminary Agreement Phase}), block size and transaction count effects prevail. On the other hand, in less congested environments (typical of the \emph{Enforcement Phase}), fee bidding reasserts itself as the prime differentiator. The live testnet environment (Sepolia)—accessed via Hardhat and Alchemy’s fee suggestions—amplifies this interplay: recommendations under EIP-1559 \cite{buterin2019eip1559} can inadvertently converge many transactions in top-tier fees, thereby reducing the relative benefit of “paying more.” This phenomenon appears clearly in Tables~\ref{table:combined_post_hoc_phase1} and \ref{table:combined_posthoc_analysis_for_phase2}, which show effect sizes that are seldom overwhelming, and in the wide whiskers of the box plots in Figures~\ref{fig:phase1-latency} and \ref{fig:low-gas-latency}.

Notably, the upper bound of gas prices observed in the \emph{Enforcement Phase} (Q5 often exceeding 100\,Gwei) is higher than in the \emph{Preliminary Agreement Phase}, where Q5 tended to stay closer to 30--35\,Gwei. We attribute this disparity partly to the dynamic base fee under EIP-1559---if the wider network's usage spiked during our Enforcement tests, recommended fees from Hardhat/Alchemy would rise accordingly. Thus, higher absolute gas-price ranges further emphasize the primacy of fee bidding once contract calls are relatively light, aligning with the moderate-to-strong effect sizes observed in Table~\ref{table:combined_posthoc_analysis_for_phase2}. Although the \emph{Enforcement Phase} typically exhibits lower median latencies due to lighter calls, a few outliers exceeded 100--140 in the box plots, likely reflecting ephemeral congestion or validator scheduling anomalies. This indicates that, despite a generally faster environment, high fees can still collide with local concurrency spikes.

From a design and operational standpoint, these observations highlight the importance of managing storage-heavy operations. When multiple large calls saturate blocks, paying only slightly higher fees may not yield faster inclusion---block propagation delays tied to big transaction sizes can overshadow fee-based prioritization \cite{croman2016scaling}. Monitoring concurrency in real time becomes essential, especially if your DApp anticipates high-volume write operations. Conversely, when transactions are lighter or sporadic, an adaptive fee strategy can ensure faster confirmations without overpaying. In practice, partial concurrency monitoring or dynamic fee estimates can help detect and react to sudden traffic spikes (even on relatively low-traffic networks like Sepolia).

These lessons bear particular relevance for inter-provider smart contracts used in multi-administrative 5G/6G networks, where multiple stakeholders (network operators, service providers, etc.) may submit updates or enforcement calls concurrently. In high-concurrency scenarios---such as when multiple providers finalize service agreements (\functionsc{addService}) or register violations (\functionsc{registerBreach}) at once---the block may fill up rapidly, causing capacity constraints that blunt the effect of high gas fees. Therefore, contract logic should minimize heavy storage writes or carefully batch complex updates during predictable off-peak windows to avoid ``traffic jams.'' Meanwhile, an adaptive fee-bidding mechanism---one that monitors real-time network conditions---can help keep costs in check while maintaining reasonable latency. Additionally, cross-domain latency requirements in 5G/6G settings make timing even more critical: if one administrative domain saturates a block with heavy calls, other providers' transactions might suffer unexpected delays, underscoring the need for concurrency-aware execution and dynamic mempool monitoring.

In closing, a holistic perspective on smart-contract development emerges from these findings. Simple heuristics---like always paying top-tier fees---cannot alone guarantee timely confirmations, particularly when ephemeral concurrency spikes or large, storage-heavy calls dominate the block. Instead, combining phase-specific awareness (heavy vs.\ light calls), real-time concurrency monitoring, and adaptive fee policies can achieve more consistent performance. For inter-provider setups in 5G/6G, these considerations become even more critical, as the complexity of coordinating multiple administrative domains raises the potential for unexpected bursts of on-chain updates. Adopting a design that partitions large storage operations, schedules bulk writes carefully, and balances fee strategies can help stakeholders avoid performance pitfalls. By integrating these insights, decentralized applications---like the one proposed in this article---can navigate the nuanced trade-offs observed in our empirical data and leverage the blockchain's fee market effectively across differing usage patterns.

\section{Conclusion}
\label{sec7}
This paper introduced a blockchain-based framework on the Ethereum live testnet to address the challenges of multi-phase inter-provider agreements in 6G contexts. By organizing our smart contracts into a \emph{Preliminary Agreement Phase} (\functionsc{addService}, \functionsc{selectService}) and an \emph{Enforcement Phase} (\functionsc{registerBreach}, \functionsc{calculatePenalty}), we measured both their gas usage and how blockchain parameters—gas price, block size, and transaction count—shape latency. Notably, our post-hoc analyses revealed that certain pairwise comparisons exhibited medium effect sizes in the \emph{Preliminary Agreement Phase} when block size or transaction count spiked, suggesting that concurrency can overshadow fee-based prioritization more strongly than initially anticipated. Meanwhile, in the \emph{Enforcement Phase}, multiple gas-price comparisons showed significant differences with a Cliff’s Delta up to 0.36, highlighting how lighter transaction logic makes fee bidding a more pivotal factor.

Such results underscore the potential benefits when designing and deploying smart contracts. For instance, minimizing heavy on-chain storage or batching it during off-peak times can alleviate block saturation, while adopting \emph{adaptive fee policies} that incorporate real-time network load can help detect and react to concurrency surges. Applying \emph{concurrency-aware scheduling} can further smooth latency spikes when multiple providers interact with the DApp. In more complex or high-throughput scenarios, additional measures such as EVM-level refactoring (e.g., reducing on-chain storage) or using Layer-2 solutions (e.g., Polygon or rollups) may mitigate latency and cost overheads. Future investigations could broaden the scope of on-chain metrics—beyond gas price, block size, and transaction count—to refine our understanding of how best to balance cost and throughput in beyond-5G decentralized environments.

%Such results underscore the need for careful calibration of contract logic and fee strategies across different phases of execution. Looking beyond basic parameters, \emph{EVM optimization techniques} (e.g., reducing on-chain storage operations) could further mitigate gas overheads for complex transactions, while concurrency-aware scheduling or batching methods might help smooth latency spikes when many users interact with the DApp simultaneously. Additionally, exploring \emph{layer-2 solutions} (e.g., Polygon or other rollups) or alternate blockchain platforms. 

%\vspace{-3mm}
\section*{Acknowledgment}
This work partially funded by Spanish MINECO grants TSI-063000-2021-54/-55 (6G-DAWN), Grant PID2021-126431OB-I00 funded by MCIN/AEI/ 10.13039/501100011033 and by “ERDF A way of making Europe” (ANEMONE), and Generalitat de Catalunya grant 2021 SGR 00770.
%\vspace{-3mm}
\balance

%\section*{REFERENCES}
\bibliographystyle{IEEEtran}
%\bibliography{./bibliography/References}

\begin{thebibliography}{10}
\providecommand{\url}[1]{#1}
\csname url@samestyle\endcsname
\providecommand{\newblock}{\relax}
\providecommand{\bibinfo}[2]{#2}
\providecommand{\BIBentrySTDinterwordspacing}{\spaceskip=0pt\relax}
\providecommand{\BIBentryALTinterwordstretchfactor}{4}
\providecommand{\BIBentryALTinterwordspacing}{\spaceskip=\fontdimen2\font plus
\BIBentryALTinterwordstretchfactor\fontdimen3\font minus \fontdimen4\font\relax}
\providecommand{\BIBforeignlanguage}[2]{{%
\expandafter\ifx\csname l@#1\endcsname\relax
\typeout{** WARNING: IEEEtran.bst: No hyphenation pattern has been}%
\typeout{** loaded for the language `#1'. Using the pattern for}%
\typeout{** the default language instead.}%
\else
\language=\csname l@#1\endcsname
\fi
#2}}
\providecommand{\BIBdecl}{\relax}
\BIBdecl

\bibitem{ETSImultidomain}
\BIBentryALTinterwordspacing
``Network functions virtualisation (nfv); management and orchestration; report on architectural options, etsi gs nfv-ifa 009 v1.1.1 (2016-07),'' 2016. [Online]. Available: \url{https://www.etsi.org/deliver/etsi_gs/nfv-ifa/001_099/009/01.01.01_60/gs_nfv-ifa009v010101p.pdf}
\BIBentrySTDinterwordspacing

\bibitem{antevski2022federation}
K.~Antevski and C.~J. Bernardos, ``Federation in dynamic environments: Can blockchain be the solution?'' \emph{IEEE Communications Magazine}, vol.~60, no.~2, pp. 32--38, 2022.

\bibitem{afraz20205g}
N.~Afraz and M.~Ruffini, ``5g network slice brokering: A distributed blockchain-based market,'' in \emph{2020 European Conference on Networks and Communications (EuCNC)}.\hskip 1em plus 0.5em minus 0.4em\relax IEEE, 2020, pp. 23--27.

\bibitem{javed2022blockchain}
F.~Javed and J.~Mangues-Bafalluy, ``Blockchain-based 6g inter-provider agreements: Auction vs. marketplace,'' in \emph{GLOBECOM 2022-2022 IEEE Global Communications Conference}.\hskip 1em plus 0.5em minus 0.4em\relax IEEE, 2022, pp. 1271--1277.

\bibitem{ETSIGSPDL024}
{European Telecommunications Standards Institute}, ``Architecture enhancements for {PDL} service provisioning in telecom networks,'' European Telecommunications Standards Institute, Sophia Antipolis, France, ETSI Group Specification ETSI GS PDL 024 V1.1.1, November 2024, available at: \url{https://www.etsi.org/deliver/etsi_gs/PDL/001_099/024/01.01.01_60/gs_PDL024v010101p.pdf}.

\bibitem{javed2022blockchainauction}
F.~Javed and J.~Mangues-Bafalluy, ``Blockchain-based 6g inter-provider agreements: Auction vs. marketplace,'' in \emph{GLOBECOM 2022-2022 IEEE Global Communications Conference}.\hskip 1em plus 0.5em minus 0.4em\relax IEEE, 2022, pp. 1271--1277.

\bibitem{javed2022blockchainPIMRC}
------, ``Blockchain and 6g networks: A use case for cost-efficient inter-provider smart contracts,'' in \emph{2022 IEEE 33rd Annual International Symposium on Personal, Indoor and Mobile Radio Communications (PIMRC)}.\hskip 1em plus 0.5em minus 0.4em\relax IEEE, 2022, pp. 602--608.

\bibitem{javed2022distributed}
F.~Javed, K.~Antevski, J.~Mangues-Bafalluy, L.~Giupponi, and C.~J. Bernardos, ``Distributed ledger technologies for network slicing: A survey,'' \emph{IEEE Access}, vol.~10, pp. 19\,412--19\,442, 2022.

\bibitem{ETSI2020Applications}
\BIBentryALTinterwordspacing
{ETSI}, ``{Permissioned Distributed Ledger (PDL); Application Scenarios},'' {European Telecommunications Standards Institute (ETSI)}, Tech. Rep. GR PDL 003 V1.1.1, Dec. 2020. [Online]. Available: \url{https://www.etsi.org/deliver/etsi_gr/PDL/001_099/003/01.01.01_60/gr_PDL003v010101p.pdf}
\BIBentrySTDinterwordspacing

\bibitem{faisal2023design}
T.~Faisal, J.~A.~O. Lucena, D.~R. Lopez, C.~Wang, and M.~Dohler, ``How to design autonomous service level agreements for 6g,'' \emph{IEEE Communications Magazine}, vol.~61, no.~3, pp. 80--85, 2023.

\bibitem{javed2024blockchain}
F.~Javed and J.~Mangues-Bafalluy, ``Blockchain-based sla management for 6g networks,'' \emph{Internet Technology Letters}, vol.~7, no.~3, p. e472, 2024.

\bibitem{ETSI2021SmartContracts}
{ETSI}, ``{Permissioned Distributed Ledgers (PDL); Smart Contracts System Architecture and Functional Specification},'' {European Telecommunications Standards Institute (ETSI)}, Tech. Rep. GR PDL 004 V1.1.1, Feb. 2021.

\bibitem{javed2023blockchain6GSLA}
F.~Javed and J.~Mangues-Bafalluy, ``Blockchain-based sla management for 6g networks,'' \emph{Internet Technology Letters}, p. e472.

\bibitem{ETSI2024Settle}
{ETSI}, ``{Permissioned Distributed Ledgers (PDL); PDL in Settlement of Usage-Based Services},'' {European Telecommunications Standards Institute (ETSI)}, Tech. Rep. GS PDL 026 V1.1.1, May 2024.

\bibitem{tm_forum_blockchain}
``Blockchain | tm forum,'' \url{https://www.tmforum.org/blockchain/}, accessed: 2023-11-06.

\bibitem{camara_blockchain}
``Blockchain public address | camara project,'' \url{https://camaraproject.org/blockchain-public-address/}, accessed: 2023-11-06.

\bibitem{etsi_pdl_017}
``Application of pdl to amended regulation 910/2014 (eidas 2) qualified trust services,'' ETSI, Technical Report ETSI GR PDL 017 V1.1.1, July 2024.

\bibitem{etsi_pdl_026}
``Pdl in settlement of usage-based services,'' ETSI, Technical Report ETSI GS PDL 026 V1.1.1, May 2024.

\bibitem{etsi_pdl_023}
``Pdl service enablers for decentralized identification and trust management,'' ETSI, Technical Report ETSI GS PDL 023 V1.1.1, April 2024.

\bibitem{faisal2022beat}
T.~Faisal, M.~Dohler, S.~Mangiante, and D.~R. Lopez, ``Beat: Blockchain-enabled accountable and transparent network sharing in 6g,'' \emph{IEEE Communications Magazine}, vol.~60, no.~4, pp. 52--56, 2022.

\bibitem{baskaran2023role}
S.~B.~M. Baskaran, T.~Faisal, C.~Wang, D.~R. Lopez, J.~Ordonez-Lucena, and I.~Arribas, ``The role of dlt for beyond 5g systems and services: A vision,'' \emph{IEEE Communications Standards Magazine}, vol.~7, no.~1, pp. 32--38, 2023.

\bibitem{zeydan2022blockchain}
E.~Zeydan, J.~Baranda, J.~Mangues-Bafalluy, Y.~Turk, and S.~B. Ozturk, ``Blockchain-based service orchestration for 5g vertical industries in multi-cloud environment,'' \emph{IEEE Transactions on Network and Service Management}, 2022.

\bibitem{augusto2024sok}
A.~Augusto, R.~Belchior, M.~Correia, A.~Vasconcelos, L.~Zhang, and T.~Hardjono, ``Sok: Security and privacy of blockchain interoperability,'' in \emph{2024 IEEE Symposium on Security and Privacy (SP)}.\hskip 1em plus 0.5em minus 0.4em\relax IEEE, 2024, pp. 3840--3865.

\bibitem{li2024security}
Y.~Li, Y.~Xiao, W.~Liang, J.~Cai, R.~Zhang, K.-C. Li, and M.~K. Khan, ``The security and privacy challenges toward cybersecurity of 6g networks: A comprehensive review,'' \emph{Computer Science and Information Systems}, no.~00, pp. 16--16, 2024.

\bibitem{hasan2024blockchain}
K.~M.~B. Hasan, M.~Sajid, M.~A. Lapina, M.~Shahid, and K.~Kotecha, ``Blockchain technology meets 6 g wireless networks: A systematic survey,'' \emph{Alexandria Engineering Journal}, vol.~92, pp. 199--220, 2024.

\bibitem{alshahrani2024enabling}
R.~Alshahrani, M.~Shabaz, M.~A. Khan, and S.~Kadry, ``Enabling intrinsic intelligence, ubiquitous learning and blockchain empowerment for trust and reliability in 6g network evolution,'' \emph{Journal of King Saud University-Computer and Information Sciences}, vol.~36, no.~4, p. 102041, 2024.

\bibitem{tripi2024security}
G.~Tripi, A.~Iacobelli, L.~Rinieri, and M.~Prandini, ``Security and trust in the 6g era: Risks and mitigations,'' \emph{Electronics}, vol.~13, no.~11, p. 2162, 2024.

\bibitem{hafi2024split}
H.~Hafi, B.~Brik, P.~A. Frangoudis, A.~Ksentini, and M.~Bagaa, ``Split federated learning for 6g enabled-networks: Requirements, challenges and future directions,'' \emph{IEEE Access}, 2024.

\bibitem{han2024lightweight}
D.~Han, Y.~Liu, R.~Cao, H.~Gao, and Y.~Lu, ``A lightweight blockchain architecture with smart collaborative and progressive evolution for privacy-preserving 6g iot,'' \emph{IEEE Wireless Communications}, 2024.

\bibitem{zahir2024performance}
A.~Zahir, M.~Groshev, K.~Antevski, C.~J.~Bernardos, C.~Ayimba, and A.~De~La~Oliva, ``Performance evaluation of private and public blockchains for multi-cloud service federation,'' in \emph{Proceedings of the 25th International Conference on Distributed Computing and Networking}, 2024, pp. 217--221.

\bibitem{antevski2023applying}
K.~Antevski and C.~J. Bernardos, ``Applying blockchain consensus mechanisms to network service federation: Analysis and performance evaluation,'' \emph{Computer Networks}, vol. 234, p. 109913, 2023.

\bibitem{wilhelmi2022end}
F.~Wilhelmi, S.~Barrachina-Mu{\~n}oz, and P.~Dini, ``End-to-end latency analysis and optimal block size of proof-of-work blockchain applications,'' \emph{IEEE Communications Letters}, vol.~26, no.~10, pp. 2332--2335, 2022.

\bibitem{afraz2023blockchain}
N.~Afraz, F.~Wilhelmi, H.~Ahmadi, and M.~Ruffini, ``Blockchain and smart contracts for telecommunications: Requirements vs. cost analysis,'' \emph{IEEE Access}, 2023.

\bibitem{ETSI2022SLA}
{ETSI}, ``{Permissioned Distributed Ledger (PDL); Reference Architecture},'' {European Telecommunications Standards Institute (ETSI)}, Tech. Rep. GS PDL 012 V1.1.1, May 2022.

\bibitem{ETSI2024Marketplace}
\BIBentryALTinterwordspacing
------, ``{Permissioned Distributed Ledgers (PDL); PDL in Wholesale Supply Chain Management},'' {European Telecommunications Standards Institute (ETSI)}, Tech. Rep. GS PDL 022 V1.1.1, Mar. 2024. [Online]. Available: \url{https://www.etsi.org/deliver/etsi_gs/PDL/001_099/022/01.01.01_60/gs_PDL022v010101p.pdf}
\BIBentrySTDinterwordspacing

\bibitem{5GAdvance}
X.~Lin, ``An overview of 5g advanced evolution in 3gpp release 18,'' \emph{IEEE Communications Standards Magazine}, vol.~6, no.~3, pp. 77--83, 2022.

\bibitem{20205growth}
C.~Papagianni, J.~Mangues-Bafalluy, P.~Bermudez, S.~Barmpounakis, D.~De~Vleeschauwer, J.~Brenes, E.~Zeydan, C.~Casetti, C.~Guimar{\~a}es, P.~Murillo \emph{et~al.}, ``{{5{G}rowth: {AI}-driven 5{G} for Automation in Vertical Industries}},'' in \emph{2020 European Conference on Networks and Communications (EuCNC)}.\hskip 1em plus 0.5em minus 0.4em\relax IEEE, 2020, pp. 17--22.

\bibitem{javed2022blockchainglobecom}
F.~Javed and J.~Mangues-Bafalluy, ``Blockchain-based 6g inter-provider agreements: Auction vs. marketplace,'' in \emph{GLOBECOM 2022-2022 IEEE Global Communications Conference}.\hskip 1em plus 0.5em minus 0.4em\relax IEEE, 2022, pp. 1271--1277.

\bibitem{ETSINFV002}
\BIBentryALTinterwordspacing
ETSI, ``{Network Functions Virtualisation {(NFV)}; Architectural Framework}.'' [Online]. Available: \url{https://www.etsi.org/}
\BIBentrySTDinterwordspacing

\bibitem{vujivcic2018blockchain}
D.~Vuji{\v{c}}i{\'c}, D.~Jagodi{\'c}, and S.~Ran{\dj}i{\'c}, ``Blockchain technology, bitcoin, and ethereum: A brief overview,'' in \emph{2018 17th international symposium infoteh-jahorina (infoteh)}.\hskip 1em plus 0.5em minus 0.4em\relax IEEE, 2018, pp. 1--6.

\bibitem{wood2014ethereum}
G.~Wood, ``Ethereum: A secure decentralised generalised transaction ledger,'' \url{https://ethereum.github.io/yellowpaper/paper.pdf}, 2014.

\bibitem{buterin2014ethereum}
\BIBentryALTinterwordspacing
V.~Buterin \emph{et~al.}, ``Ethereum: A secure decentralized transaction ledger,'' 2014, accessed: 2024-09-17. [Online]. Available: \url{https://ethereum.org/en/developers/docs/gas/}
\BIBentrySTDinterwordspacing

\bibitem{EthereumSmartContracts}
\BIBentryALTinterwordspacing
``Smart contracts,'' Ethereum Developers Documentation, accessed: 2025-01-06. [Online]. Available: \url{https://ethereum.org/en/developers/docs/smart-contracts/}
\BIBentrySTDinterwordspacing

\bibitem{croman2016scaling}
\BIBentryALTinterwordspacing
K.~Croman, I.~Eyal, and et~al., ``On scaling decentralized blockchains,'' in \emph{Conference on Financial Cryptography and Data Security}.\hskip 1em plus 0.5em minus 0.4em\relax Springer, Berlin, Heidelberg, 2016, accessed: [Insert Access Date]. [Online]. Available: \url{https://eprint.iacr.org/2016/199.pdf}
\BIBentrySTDinterwordspacing

\bibitem{kiayias2017ouroboros}
A.~Kiayias, A.~Russell, B.~David, and R.~Oliynykov, ``Ouroboros: A provably secure proof-of-stake blockchain protocol,'' in \emph{Annual International Cryptology Conference (CRYPTO)}, 2017, pp. 357--388, foundational work on PoS security and performance, relevant to Ethereum’s switch from PoW to PoS.

\bibitem{eyal2016bitcoinng}
I.~Eyal, A.~E. Gencer, E.~G. Sirer, and R.~van Renesse, ``Bitcoin-ng: A scalable blockchain protocol,'' in \emph{Proceedings of the 13th USENIX Symposium on Networked Systems Design and Implementation (NSDI)}, 2016, pp. 45--59, although focused on Bitcoin, it discusses scalability and consensus issues broadly relevant to blockchain designs.

\bibitem{EthereumTransactions}
\BIBentryALTinterwordspacing
``Transactions,'' Ethereum Developers Documentation, accessed: 2025-01-06. [Online]. Available: \url{https://ethereum.org/en/developers/docs/transactions/}
\BIBentrySTDinterwordspacing

\bibitem{EthereumProofOfStake}
\BIBentryALTinterwordspacing
``Proof of stake (pos),'' Ethereum Developers Documentation, accessed: 2025-01-06. [Online]. Available: \url{https://ethereum.org/en/developers/docs/consensus-mechanisms/pos/}
\BIBentrySTDinterwordspacing

\bibitem{sai2021performance}
K.~Sai, Y.~Yu, M.~Stabauer, and P.~Felber, ``Performance evaluation of permissioned ethereum-based blockchains,'' \emph{Future Generation Computer Systems}, vol. 124, pp. 12--22, 2021, empirical analysis of Ethereum variants, focusing on performance metrics such as latency.

\bibitem{EthereumVirtualMachine}
\BIBentryALTinterwordspacing
``Ethereum virtual machine (evm),'' Ethereum Developers Documentation, accessed: 2025-01-06. [Online]. Available: \url{https://ethereum.org/en/developers/docs/evm/}
\BIBentrySTDinterwordspacing

\bibitem{chen2017contracts}
T.~Chen, X.~Li, X.~Luo, and T.~Zhang, ``Under-optimized smart contracts devour your money,'' in \emph{Proceedings of the 24th IEEE International Conference on Software Analysis, Evolution and Reengineering (SANER)}, 2017, pp. 442--446, explores gas inefficiencies in contracts, tying in well with gas usage and cost discussions.

\bibitem{bonneau2015sok}
J.~Bonneau, A.~Miller, J.~Clark, A.~Narayanan, J.~A. Kroll, and E.~W. Felten, ``Sok: Research perspectives and challenges for bitcoin and cryptocurrencies,'' in \emph{2015 IEEE Symposium on Security and Privacy}, 2015, pp. 104--121, systematizes knowledge on cryptocurrencies, discussing many aspects that impact latency and throughput.

\bibitem{androulaki2018hyperledger}
E.~Androulaki, A.~Barger, V.~Bortnikov, S.~Muralidharan, C.~Cachin, K.~Christidis, A.~D. Caro, E.~Androulaki, and M.~Vukolic, ``Hyperledger fabric: A distributed operating system for permissioned blockchains,'' in \emph{Proceedings of the 13th EuroSys Conference}, 2018, pp. 1--15, focuses on permissioned blockchains but includes many concepts applicable to throughput, latency, etc.

\bibitem{decentralizedstorage2024}
\BIBentryALTinterwordspacing
minimalsm, ``Decentralized storage,'' September 2024, accessed: 2025-02-09. [Online]. Available: \url{https://ethereum.org/en/developers/docs/storage/decentralized-storage}
\BIBentrySTDinterwordspacing

\bibitem{EtherVM}
\BIBentryALTinterwordspacing
EtherVM, ``Ethervm - ethereum virtual machine resources,'' 2025, accessed: 2025-02-15. [Online]. Available: \url{https://ethervm.io/}
\BIBentrySTDinterwordspacing

\bibitem{eyal2014majority}
I.~Eyal and E.~G. Sirer, ``Majority is not enough: Bitcoin mining is vulnerable,'' \url{https://eprint.iacr.org/2014/718.pdf}, 2014, accessed: [Insert Access Date].

\bibitem{decker2013information}
\BIBentryALTinterwordspacing
C.~Decker and R.~Wattenhofer, ``Information propagation in the bitcoin network,'' \emph{arXiv preprint arXiv:1304.1100}, 2013, accessed: [Insert Access Date]. [Online]. Available: \url{https://arxiv.org/abs/1304.1100}
\BIBentrySTDinterwordspacing

\bibitem{luu2015incentives}
L.~Luu, J.~Teutsch, R.~Kulkarni, and P.~Saxena, ``Demystifying incentives in the consensus computer,'' in \emph{Proceedings of the 22nd ACM SIGSAC Conference on Computer and Communications Security (CCS)}, 2015, pp. 706--719, investigates how incentives (gas price, fees) drive network behavior in blockchains like Ethereum.

\bibitem{buterin2019eip1559}
V.~Buterin, ``Eip-1559: Fee market change for eth 1.0 chain,'' \url{https://eips.ethereum.org/EIPS/eip-1559}, 2019, accessed: 2024-04-27.

\bibitem{gervais2016security}
A.~Gervais and et~al., ``On the security and performance of proof of work blockchains,'' \url{https://eprint.iacr.org/2016/999.pdf}, 2016, accessed: [Insert Access Date].

\bibitem{monrat2019survey}
A.~A. Monrat, O.~Schelén, and K.~Andersson, ``A survey of blockchain from the perspectives of applications, challenges, and opportunities,'' \emph{IEEE Access}, vol.~7, pp. 117\,134--117\,151, 2019, broad survey covering blockchain performance, scalability, and use cases.

\end{thebibliography}

\end{document}